\documentclass[aps,prl,reprint,nofootinbib]{revtex4-2}
\usepackage{epsfig}
\usepackage{amsmath,amssymb,amsfonts}
\usepackage{mathrsfs}
\usepackage{bbm}
\usepackage{slashed}
\usepackage{graphicx}
\usepackage{verbatim}
\usepackage[T1]{fontenc}
\usepackage[utf8]{inputenc}
\usepackage{hyperref}
\usepackage{ifthen}
\usepackage{forloop}
\usepackage[english]{babel}
\usepackage{mathbbol}
\usepackage[usenames]{xcolor}
\usepackage[normalem]{ulem}
\usepackage{dcolumn}
\usepackage{rotating}
\usepackage{longtable}

\definecolor{bluecite}{HTML}{0875b7}

\hypersetup{
	colorlinks=true,
	linkcolor=bluecite,
	urlcolor=magenta,
	citecolor=bluecite
}

\newcommand{\GR}{{\small GR}}
\newcommand{\eg}{{\textit{e.g.}}}

\newcommand{\precision}{256}

\allowdisplaybreaks

\begin{document}

\title{One-loop renormalisation of cubic gravity in six dimensions}
\author{Benjamin Knorr\,\href{https://orcid.org/0000-0001-6700-6501}{\protect \includegraphics[scale=.07]{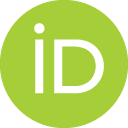}}}
\email[]{bknorr@perimeterinstitute.ca}
\affiliation{
Perimeter Institute for Theoretical Physics, 31 Caroline Street North, Waterloo, ON N2L 2Y5, Canada
}

\begin{abstract}
We present the complete set of universal one-loop beta functions of cubic gravity in six dimensions. The system admits over 8000 
distinct fixed points, of which more than 200 are real. Some of them might be relevant for the quantisation of gravity in the physical case of four dimensions.
\end{abstract}

\maketitle

\textit{Introduction.} --- One of the major open problems in theoretical physics is to find a theory of quantum gravity. Such a theory is needed, among other reasons, to understand the fate of spacetime singularities predicted by General Relativity (\GR{}) for example at the inside of black holes or at the big bang \cite{Hawking:1969sw}. While a lot of different paths have been investigated, no single approach can claim complete success yet.

The fundamental roadblock in the quantisation of \GR{} in four dimensions is that Newton's constant is not asymptotically free, so that the perturbative renormalisation program, which works extremely well for the Standard Model, is doomed to fail in gravity.\footnote{See however \cite{Morris:2018upm, Morris:2018axr, Mitchell:2020fjy, Kellett:2020mle} for an approach trying to circumvent this problem in a non-standard way.} While effective field theory can be used to systematically compute quantum corrections \cite{Donoghue:1993eb, Donoghue:1994dn}, this is limited to energies below the Planck energy, or equivalently to distances larger than the Planck length. A way around this problem, while staying within the perturbative paradigm, is to consider higher derivative theories. The prime example is Stelle gravity \cite{Stelle:1976gc, Stelle:1977ry}, which is asymptotically free in four dimensions. While classically such theories suffer from the Ostrogradsky instability \cite{Ostrogradsky:1850fid, Woodard:2015zca}, their quantum versions might be well-defined by themselves \cite{Donoghue:2019ecz, Donoghue:2021eto}, or by non-perturbative effects which remove the additional poles \cite{Platania:2020knd}.

From the point of view of critical phenomena, considering a given higher derivative action corresponds to choosing the upper critical dimension of the theory. It is for example well-known that the upper critical dimension of the Ising universality class is $d=4$. In particular, the Wilson-Fisher fixed point associated to the second order phase transition persists upon lowering the dimension \cite{Wilson:1971dc, Codello:2012sc}. As a matter of fact, perturbative methods like the $(4-\epsilon)$- and $(2+\epsilon)$-expansions strictly rely on the existence of a continuous connection of non-trivial fixed points to the corresponding Gaussian fixed point.

This view, applied to gravity, was recently advocated in \cite{Martini:2021lcx}. Therein, the authors suggest to consider cubic gravity in six dimensions, where all of its coupling constants are dimensionless, and to study whether a suitable fixed point exists which extends all the way down to the physical dimension $d=4$. Cubic gravity seems to be the most interesting case beyond Stelle gravity, since its critical dimension is closest to the physical dimension. Its action also features the well-known Goroff-Sagnotti term, which appears as the first non-subtractable divergence when renormalising \GR{} at the two-loop order in $d=4$ \cite{Goroff:1985sz, Goroff:1985th, vandeVen:1991gw}. For a non-perturbative treatment of this term in $d=4$, see \cite{Gies:2016con}. Similar ideas have been discussed earlier from the ``opposite'' direction, starting with \GR{} in $d=2$, see \eg{} \cite{Kawai:1989yh, Jack:1990ey, Kawai:1992np, Kawai:1993mb, Aida:1994np, Aida:1996zn}, and \cite{Martini:2021slj} for a modern perspective.

Partial results at the cubic order on a particular background have been obtained previously in \cite{Pang:2012rd}. For a recent computation of some beta functions within cubic gravity in $d=4$ see \cite{Rachwal:2021bgb}. The conformally reduced version has been investigated non-perturbatively in $d=4$ in \cite{Knorr:2020ckv}. Results on a particular subclass of (classical) cubic gravity theories can be found in \eg{} \cite{Bueno:2016xff, Hennigar:2016gkm, Bueno:2016lrh, Bueno:2016ypa, Hennigar:2018hza, Jimenez:2020gbw}.

The present Letter will take a major step in the investigation of cubic quantum gravity by computing and analysing the complete set of universal one-loop beta functions in six dimensions. We find a very intricate phase diagram with thousands of fixed points. If this structure persists down to the physical dimension, it gives a whole new perspective on the phase diagram of quantum gravity.

\textit{Setup.} --- We start by fixing our conventions for the (Euclidean) action. It reads
\begin{widetext}
\begin{equation}\label{eq:action}
\begin{aligned}
 &S = \frac{1}{\lambda_6} \int \text{d}^6x \, \sqrt{g} \Bigg[ \frac{1}{3} C^{\mu\nu\rho\sigma} \, \Delta \, C_{\mu\nu\rho\sigma} - \frac{\omega_6}{5} R \, \Delta \, R + \rho_{R^3} R^3 + \rho_{RS^2} R \, S^{\mu\nu} S_{\mu\nu} + \rho_{RC^2} R \, C^{\mu\nu\rho\sigma} C_{\mu\nu\rho\sigma} + \rho_{S^3} S^\mu_{\phantom{\mu}\nu} S^\nu_{\phantom{\nu}\rho} S^\rho_{\phantom{\rho}\mu} \\
 & \hspace{5cm} + \rho_{S^2C} S^{\mu\nu} S^{\rho\sigma} C_{\mu\rho\nu\sigma} + \rho_{SC^2} S^{\mu\nu} C_{\mu\alpha\beta\gamma} C_\nu^{\phantom{\nu}\alpha\beta\gamma} + \rho_{C^3} C^{\mu\nu}_{\phantom{\mu\nu}\rho\sigma} C^{\rho\sigma}_{\phantom{\rho\sigma}\tau\omega} C^{\tau\omega}_{\phantom{\tau\omega}\mu\nu} - \frac{1}{8} \rho_{\mathfrak E} \mathfrak E_6 \bigg] \, .
\end{aligned}
\end{equation}
\end{widetext}
In this, $C$ is the Weyl tensor, $S$ is the trace-free Ricci tensor, $R$ is the Ricci scalar and $\Delta=-D^2$ is the covariant Laplacian of the metric $g$. The factors $\lambda_6, \omega_6$ and $\rho_X$ are dimensionless coupling constants. The main goal of this Letter is to compute and analyse their beta functions.

Our conventions for the numerical prefactors of the monomials $C\Delta C$ and $R\Delta R$ normalise the kinetic terms of the spin two and zero modes. In particular, for positive couplings $\lambda_6, \omega_6 > 0$, both modes have the same sign for their propagators as in \GR{}. That is, the propagator has a positive sign for the spin two mode, and a negative sign for the conformal mode. The last term indicates the topological Euler term in $d=6$, given by
\begin{widetext}
\begin{equation}\label{eq:Euler}
\begin{aligned}
\mathfrak E_6 &= R^3 - 12 R R^{\mu\nu} R_{\mu\nu} + 3 R R^{\mu\nu\rho\sigma} R_{\mu\nu\rho\sigma} + 16 R^\mu_{\phantom{\mu}\nu} R^\nu_{\phantom{\nu}\rho} R^\rho_{\phantom{\rho}\mu} + 24 R^{\mu\nu} R^{\rho\sigma} R_{\mu\rho\nu\sigma} - 24 R^{\mu\nu} R_{\mu\alpha\beta\gamma} R_\nu^{\phantom{\nu}\alpha\beta\gamma} \\
 & \qquad+ 4 R^{\mu\nu}_{\phantom{\mu\nu}\rho\sigma} R^{\rho\sigma}_{\phantom{\rho\sigma}\tau\omega} R^{\tau\omega}_{\phantom{\tau\omega}\mu\nu} - 8 R_{\mu\phantom{\rho}\nu}^{\phantom{\mu}\rho\phantom{\nu}\sigma} R_{\rho\phantom{\tau}\sigma}^{\phantom{\rho}\tau\phantom{\sigma}\omega} R_{\tau\phantom{\mu}\omega}^{\phantom{\tau}\mu\phantom{\omega}\nu} \, ,
\end{aligned}
\end{equation}
\end{widetext}
and its prefactor is normalised to one of the combinations cubic in the Riemann tensor. We do not translate the Euler term into the traceless basis to avoid lengthy dimension-dependent prefactors. Due to its topological nature, its coupling constant $\rho_{\mathfrak E}$ will only appear in its own beta function. The monomial with coupling $\rho_{SC^2}$, and the six-dimensional Euler term vanish in $d=4$. The set of curvature monomials is complete in $d=6$ up to boundary terms which we shall neglect \cite{Fulling:1992vm}. In particular, the relation
\begin{widetext}
\begin{equation}
\begin{aligned}
 S^{\mu\nu} \Delta S_{\mu\nu} &\simeq \frac{1}{3} C^{\mu\nu\rho\sigma} \Delta C_{\mu\nu\rho\sigma} + \frac{2}{15} R \Delta R - \frac{1}{5} R S^{\mu\nu} S_{\mu\nu} + \frac{1}{9} R C^{\mu\nu\rho\sigma} C_{\mu\nu\rho\sigma} - \frac{3}{2} S^\mu_{\phantom{\mu}\nu} S^\nu_{\phantom{\nu}\rho} S^\rho_{\phantom{\rho}\mu} \\
 & \quad + S^{\mu\nu} S^{\rho\sigma} C_{\mu\rho\nu\sigma} + \frac{2}{3} S^{\mu\nu} C_{\mu\alpha\beta\gamma} C_\nu^{\phantom{\nu}\alpha\beta\gamma} - \frac{1}{3} C^{\mu\nu}_{\phantom{\mu\nu}\rho\sigma} C^{\rho\sigma}_{\phantom{\rho\sigma}\tau\omega} C^{\tau\omega}_{\phantom{\tau\omega}\mu\nu} - \frac{4}{3} C_{\mu\phantom{\rho}\nu}^{\phantom{\mu}\rho\phantom{\nu}\sigma} C_{\rho\phantom{\tau}\sigma}^{\phantom{\rho}\tau\phantom{\sigma}\omega} C_{\tau\phantom{\mu}\omega}^{\phantom{\tau}\mu\phantom{\omega}\nu} \, ,
\end{aligned}
\end{equation}
\end{widetext}
is valid up to total derivatives and in $d=6$.

Due to its invariance under diffeomorphisms, the action \eqref{eq:action} has to be supplemented by a gauge fixing. In this Letter we mimic the successful strategy previously employed in the computation of the one-loop beta functions in Stelle gravity, see \eg{} \cite{deBerredoPeixoto:2004if, Groh:2011vn, Falls:2020qhj}. We use the background field method, splitting the full metric via
\begin{equation}
 g_{\mu\nu} = \bar g_{\mu\nu} + h_{\mu\nu} \, ,
\end{equation}
and choose a gauge fixing which brings the flat part of the two-point function into minimal form. This allows for a straightforward derivative expansion of the functional one-loop trace. It can be achieved by supplementing the action with a gauge fixing term of the form
\begin{equation}\label{eq:Sgf}
 S^\text{gf} = \frac{1}{2\lambda_6} \int \text{d}^6x \, \sqrt{\bar g} \, \mathcal F^\mu \left[ \mathcal Y^2 \right]_\mu^{\phantom{\mu}\nu} \mathcal F_\nu \, .
\end{equation}
Here,
\begin{equation}\label{eq:Fgf}
 \mathcal F_\mu = \bar D^\alpha h_{\mu\alpha} - \frac{1+4 \omega_6}{6+4\omega_6} \bar D_\mu h^\alpha_{\phantom{\alpha}\alpha} \, ,
\end{equation}
is a linear covariant gauge condition, and
\begin{equation}\label{eq:Ygf}
 \mathcal Y_\mu^{\phantom{\mu}\nu} = \bar \Delta \delta_\mu^{\phantom{\mu}\nu} - \sqrt{\frac{3+2\omega_6}{5}} \, \bar D_\mu \bar D^\nu + \bar D^\nu \bar D_\mu \, ,
\end{equation}
is an additional weight function. The gauge fixing gives rise to standard ghost contributions. Choosing the square of $\mathcal Y$ in \eqref{eq:Sgf} instead of a fourth order operator directly simplifies the computation of the corresponding ``third'' ghost trace, since
\begin{equation}\label{eq:trY}
 \text{tr} \ln \left[ \mathcal Y^2 \right] = 2 \, \text{tr} \ln \mathcal Y \, .
\end{equation}
This construction generalises straightforwardly to higher order gravity theories, where only the power of $\mathcal Y$ in \eqref{eq:Sgf} and the non-trivial coefficients in \eqref{eq:Fgf} and \eqref{eq:Ygf} have to be adapted accordingly.

Note that the appearance of the square root of the coupling $\omega_6$ in \eqref{eq:Ygf} is spurious: taking the square of $\mathcal Y$ gives
\begin{equation}\label{eq:Ygf2}
 \left[ \mathcal Y^2 \right]_\mu^{\phantom{\mu}\nu} = \left[ \left( \bar \Delta \mathbbm 1+ \overline{\text{Ric}} \right)^2 \right]_\mu^{\phantom{\mu}\nu} - \frac{2}{5}(\omega_6-1) \bar D_\mu \bar \Delta \bar D^\nu \, .
\end{equation}
Here $\overline{\text{Ric}}$ indicates the background Ricci tensor. We can see that the square root drops out. Moreover, for the trace of the ``third'' ghost \eqref{eq:trY}, we observe that $\mathcal Y$ itself decomposes into transverse and longitudinal parts which do not mix, so that this part of the one-loop trace is completely independent of $\omega_6$. The appearance of $\omega_6$ in denominators as in \eqref{eq:Fgf} is however not cancelled in \eqref{eq:Sgf} directly, and serves as a check for the correctness of the computation, since the final beta functions must be polynomial in all couplings (except for isolated negative powers of $\omega_6$ which come from the spin zero propagator). This structure carries over to higher order theories.

As an additional check for the correctness of our result, we added a term $\kappa \bar R_\mu^{\phantom{\mu}\nu}$ to the operator \eqref{eq:Ygf}, and verified that the final beta functions do not depend on $\kappa$, indicating the gauge independence of the result.

The practical computation of the one-loop traces was carried out with standard heat kernel techniques \cite{Vassilevich:2003xt, Groh:2011dw} and the functional renormalisation group \cite{Wetterich:1992yh, Ellwanger:1993mw, Morris:1993qb, Reuter:1996cp, Berges:2000ew, Pawlowski:2005xe, Gies:2006wv, Berges:2012ty, Percacci:2017fkn, Reuter:2019byg, Dupuis:2020fhh, Reichert:2020mja}. The regularisation follows refs. \cite{Groh:2011vn, Falls:2020qhj} in the graviton sector and ref. \cite{Knorr:2021slg} in the ghost sector, but the final result is expected to be independent of the concrete choice. All appearing threshold integrals are computable without specifying a concrete regulator shape, which follows from general structural considerations, and thus does not constitute an extra check. The algebra was handled with the Mathematica package \emph{xAct} \cite{Brizuela:2008ra, 2007CoPhC.177..640M, 2008CoPhC.179..586M, 2008CoPhC.179..597M, 2014CoPhC.185.1719N}, and \cite{xParallel} was used to parallelise the code. The code was tested extensively and implements all the optimisations discussed in \cite{Knorr:2021slg}.

\textit{Beta functions.} --- The full set of beta functions for all couplings appearing in \eqref{eq:action} is too long to be displayed in the main text. They are presented in full in the supplemental text, and are also contained in a supplemental Mathematica notebook for easy accessibility. The beta functions constitute the first main result of this work. For illustration, here we present reduced beta functions where we have set all couplings except $\lambda_6$ and $\omega_6$ to zero. They read:
\begin{subequations}\label{eq:betafuncs}
\begin{align}
 \beta_{\lambda_6} &= \frac{1}{(4\pi)^3} \frac{\lambda_6^2}{\omega_6} \left[ \frac{7}{80} \left( 1 - \frac{15784}{147} \omega_6 \right) \right] , \\
 \beta_{\omega_6} &= \frac{1}{(4\pi)^3} \lambda_6 \left[ \frac{155}{16} \left( 1 - \frac{200}{217} \omega_6 \right) \right] , \\
 \beta_{\rho_{R^3}} &= \frac{1}{(4\pi)^3} \frac{\lambda_6}{\omega_6} \left[ \frac{77}{2430000} \left( 1 - \frac{45107}{42} \omega_6 \right) \right] , \\
 \beta_{\rho_{RS^2}} &= \frac{1}{(4\pi)^3} \frac{\lambda_6}{\omega_6} \left[ -\frac{77}{108000} \left( 1 - \frac{572857}{462} \omega_6 \right) \right] , \\
 \beta_{\rho_{RC^2}} &= \frac{1}{(4\pi)^3} \frac{\lambda_6}{\omega_6} \left[ \frac{16967}{324000} \left( 1 - \frac{53224519}{712614} \omega_6 \right) \right] , \\
 \beta_{\rho_{S^3}} &= \frac{1}{(4\pi)^3} \frac{\lambda_6}{\omega_6} \left[ \frac{77}{43200} \left( 1 + \frac{48574}{231} \omega_6 \right) \right] , \\
 \beta_{\rho_{S^2C}} &= \frac{1}{(4\pi)^3} \frac{\lambda_6}{\omega_6} \left[ \frac{77}{21600} \left( 1 - \frac{76616}{231} \omega_6 \right) \right] , \\
 \beta_{\rho_{SC^2}} &= \frac{1}{(4\pi)^3} \frac{\lambda_6}{\omega_6} \left[ -\frac{623}{10800} \left( 1 + \frac{1472452}{13083} \omega_6 \right) \right] , \\
 \beta_{\rho_{C^3}} &= \frac{1}{(4\pi)^3} \frac{\lambda_6}{\omega_6} \left[ \frac{97}{10800} \left( 1 + \frac{2530898}{2037} \omega_6 \right) \right] , \\
 \beta_{\rho_{\mathfrak E}} &= \frac{1}{(4\pi)^3} \frac{\lambda_6}{\omega_6} \left[ \frac{77}{8100} \left( 1 + \frac{5831501}{3234} \omega_6 \right) \right] .
\end{align}
\end{subequations}
Interestingly, all truncated equations consist of a sum of two terms only. By contrast, in quadratic gravity in $d=4$, the beta function of the coupling of the squared Ricci scalar is quadratic in itself.

The full beta functions vanish for $\lambda_6=0$, indicating the Gaussian fixed point. As expected from general grounds, the beta functions are polynomials of up to cubic order in the couplings $\rho_X$, and contain both positive and negative powers of $\omega_6$. Some of the couplings only contribute with certain powers, or not at all, to some of the beta functions. This comes from the fact that the beta functions arise from the trace of the second variation of the action. For example, the coupling $\rho_{R^3}$ cannot contribute to any beta function belonging to a tensor structure which does not contain at least one power of the Ricci scalar. A notable difference to quadratic gravity is that the beta function of the coupling which controls the physical spin two propagator, $\lambda_6$, also depends on the other couplings.

\textit{Fixed points.} --- We now discuss the (non-trivial) fixed point structure to identify different universality classes. A non-trivial fixed point is achieved at finite values of the couplings except $\lambda_6$ so that all beta functions except $\beta_{\lambda_6}$ vanish, and complementing this with $\lambda_6\to0$ to also impose $\beta_{\lambda_6}=0$. This corresponds to a rescaling of the renormalisation group ``time'' by $\lambda_6$ \cite{Martini:2021lcx}. Note that since the one-loop beta functions are expected to be gauge-independent, our specific choice of gauge fixing, \eqref{eq:Fgf} and \eqref{eq:Ygf}, does not impose any restrictions on the coupling $\omega_6$. In particular, $\omega_6 \leq -\tfrac{3}{2}$ is allowed at the level of the beta functions. Finally, since only $\beta_{\rho_{\mathfrak E}}$ depends on $\rho_{\mathfrak E}$, we can use it to unambiguously determine the fixed point value of $\rho_{\mathfrak E}$.

We are thus left with the task to find the roots of eight polynomials in eight variables. From this general structure alone, it is clear that cubic gravity allows for a very rich phase diagram, potentially with a very large number of different phases. To solve this system of equations, ideally we would construct a Gr\"obner basis \cite{Buchberger2010} to analytically prove that we have found all solutions. For truncations of the system, this indeed works. For example, taking into account the couplings $\lambda_6, \omega_6, \rho_{R^3}, \rho_{RS^2}, \rho_{RC^2}$ and $\rho_{S^3}$, we were able to construct the complete Gr\"obner basis. The fixed points of this reduced system are related to the roots of a polynomial of order 363, with integer coefficients with absolute values between $10^{595}$ and $10^{1023}$. The system is thus extremely difficult to treat, both analytically and numerically. Unfortunately, we were not able to compute the Gr\"obner basis for the complete system analytically. In the following, we thus resort to arbitrary precision numerical solving methods with the NSolve routine of Mathematica.\footnote{In practice, we used NSolve at machine precision and used the result as a starting point for FindRoot with higher precision. Using NSolve directly at higher precision did not terminate after a five week runtime.} From experimenting in truncations where we have access to the analytical Gr\"obner basis, we see that not all fixed points are found if we use a precision which is too low. Since it is difficult to know from the numerical results alone when the precision is high enough to discover all fixed points, all numbers below are lower bounds on the true number of fixed points.

With this numerical setup, and its limitations in mind, we have found 8044 fixed points, of which 220 are real, and the remaining 7824 are pairs of complex conjugate solutions. These fixed points have been computed with \precision{} digits of precision, and it has been verified that all of them satisfy the fixed point condition with only a few digits of precision lost. A complete list of the real fixed points that we have found is given in the supplemental text, and the full list of 8044 fixed points is included in the same supplemental Mathematica notebook that also includes the beta functions. The set of fixed points constitutes the second main result of this work.

We find that the total number of fixed points is even. From the study of condensed matter systems in dependence on the dimension \cite{Eichhorn:2015woa}, we know that fixed point creation and annihilation processes typically involve pairs of fixed points. This indicates that in the physical dimension $d=4$, we expect an even number of non-trivial fixed points which are connected to the universality classes characterised by cubic gravity. Note that not all of them need to lie in the physical regime - for example, for some of those fixed points, Newton's constant could be negative, and the corresponding fixed point can be discarded on those grounds. Such additional constraints would allow for the possibility of a unique physical fixed point in $d=4$. It is however conceivable that an odd number of real fixed points could have been missed due to the limitations of our numerical search. In that case one would expect an odd number of real fixed points related to cubic gravity in $d=4$.

\textit{Outlook.} --- The derivation of the full set of one-loop beta functions is the first major step in the investigation of whether the corresponding universality classes of cubic gravity are relevant for the renormalisation of gravity in four dimensions, and thus for the real world. We find a plethora of fixed points, but likely not all of them will play a role in $d=4$. This is in stark contrast to Stelle gravity, which only features two fixed points. It would be very interesting to study the phase diagram as a function of the dimension. This has particular relevance for the asymptotic safety programme, especially since so far, the full non-perturbative beta functions of the cubic terms have not been resolved (for partial results, see \eg{} \cite{Gies:2016con, Falls:2017lst, Falls:2018ylp, deBrito:2020xhy, Kluth:2020bdv}). Extrapolating from our results, one might find a much richer phase structure than previously expected at this order also in $d=4$.

A potential additional selection criterion for viable fixed points could come from conditions on the sign of some of the couplings to ensure unitarity and causality of the theory \cite{Adams:2006sv, Donoghue:2021meq} or compatibility with swampland conjectures \cite{Basile:2021krr}. Clearly, higher loop computations, or non-perturbative computations in $d=4$ will be necessary to give a reliable answer. We will report on the non-perturbative renormalisation of cubic gravity in $d=4$ elsewhere. Finally, an open question is the relevance of the beta functions corresponding to operators that vanish in $d=4$, and how the contribution of these couplings to the renormalisation group flow decouples below the dimension where the corresponding operator vanishes. We leave this intriguing question for future work.

\bigskip

\acknowledgments

\textit{Acknowledgments.} --- I would like to thank Alessia Platania for helpful comments on the manuscript. This work was supported by Perimeter Institute for Theoretical Physics. Research at Perimeter Institute is supported in part by the Government of Canada through the Department of Innovation, Science and Economic Development and by the Province of Ontario through the Ministry of Colleges and Universities.

\bibliographystyle{apsrev4-2}
\bibliography{general_bib}

\widetext

\renewcommand{\theequation}{A\arabic{equation}}
\setcounter{equation}{0}

\begin{center}
	\textbf{\large Supplemental Material} 
\end{center}
\section{A - Complete beta functions}
Here we present the complete set of beta functions for cubic gravity in $d=6$. The full trace has the form
\begin{equation}
\begin{aligned}
	\mathcal T = \int \text{d}^6x \, \sqrt{g} \Bigg[ &t_1 C^{\mu\nu\rho\sigma} \, \Delta \, C_{\mu\nu\rho\sigma} + t_2 R \, \Delta \, R + t_3 R^3 + t_4 R \, S^{\mu\nu} S_{\mu\nu} + t_5 R \, C^{\mu\nu\rho\sigma} C_{\mu\nu\rho\sigma} + t_6 S^\mu_{\phantom{\mu}\nu} S^\nu_{\phantom{\nu}\rho} S^\rho_{\phantom{\rho}\mu} \\
	&\qquad\qquad\qquad + t_7 S^{\mu\nu} S^{\rho\sigma} C_{\mu\rho\nu\sigma} + t_8 S^{\mu\nu} C_{\mu\alpha\beta\gamma} C_\nu^{\phantom{\nu}\alpha\beta\gamma} + t_9 C^{\mu\nu}_{\phantom{\mu\nu}\rho\sigma} C^{\rho\sigma}_{\phantom{\rho\sigma}\tau\omega} C^{\tau\omega}_{\phantom{\tau\omega}\mu\nu} + t_{10} \rho_{\mathfrak E} \mathfrak E_6 \bigg] \, ,
\end{aligned}
\end{equation}
where the $t_i$ are given below. To obtain the beta functions, one has to compare the coefficients of the tensor monomials with the scale derivative of the action \eqref{eq:action}, that is
\begin{equation}
	\sum_{g_i} \beta_{g_i} \frac{\partial}{\partial g_i} S = \mathcal T \, ,
\end{equation}
where the sum extends over all 10 couplings, $g_i \in \{ \lambda_6, \omega_6, \rho_{R^3}, \rho_{RS^2}, \rho_{RC^2}, \rho_{S^3}, \rho_{S^2C}, \rho_{SC^2}, \rho_{C^3}, \rho_{\mathfrak E} \}$. The beta functions are thus related to the coefficients $t_i$ via
\begin{subequations}
\begin{align}
	\beta_{\lambda_6} &= - 3 \lambda_6^2 t_1 \, , \\
	\beta_{\omega_6} &= -\lambda_6 \left( 3\omega_6 t_1 + 5 t_2 \right) \, , \\
	\beta_{\rho_{R^3}} &= \lambda_6 \left( -3\rho_{R^3} t_1 + t_3 \right) \, , \\
	\beta_{\rho_{RS^2}} &= \lambda_6 \left( -3\rho_{RS^2} t_1 + t_4 \right) \, , \\
	\beta_{\rho_{RC^2}} &= \lambda_6 \left( -3\rho_{RC^2} t_1 + t_5 \right) \, , \\
	\beta_{\rho_{S^3}} &= \lambda_6 \left( -3\rho_{S^3} t_1 + t_6 \right) \, , \\
	\beta_{\rho_{S^2C}} &= \lambda_6 \left( -3\rho_{S^2C} t_1 + t_7 \right) \, , \\
	\beta_{\rho_{SC^2}} &= \lambda_6 \left( -3\rho_{SC^2} t_1 + t_8 \right) \, , \\
	\beta_{\rho_{C^3}} &= \lambda_6 \left( -3\rho_{C^3} t_1 + t_9 \right) \, , \\
	\beta_{\rho_{\mathfrak E}} &= -\lambda_6 \left( 3 \rho_{\mathfrak E} t_1 + 8 t_{10} \right) \, .
\end{align}
\end{subequations}
The coefficients $t_i$ are collected in the subsequent equations.

\clearpage

\begin{sideways}
	\begin{minipage}{\textheight}
		\begin{align}
			t_1 &= \frac{1}{(4\pi)^3} \frac{1}{\omega_6^2} \Bigg[ \omega_6^2 \bigg\{-\frac{87 \rho_{C^3}}{160}+\rho_{C S^2} \left(\frac{9 \rho_{C^3}}{32}-\frac{123 \rho_{\text{SC}^2}}{320}-\frac{251}{400}\right)+\rho_{S^3} \left(-\frac{351\rho_{C^3}}{160}-\frac{837 \rho_{C S^2}}{2000}-\frac{69 \rho_{\text{SC}^2}}{80}-\frac{857}{600}\right) \notag \\
			&\hspace{3.25cm} +\left(-\frac{261 \rho_{C^3}}{640}-\frac{493}{480}\right) \rho_{\text{SC}^2}+\frac{3 \rho_{C S^2}^2}{2000}-\frac{147 \rho_{S^3}^2}{500}-\frac{27 \rho_{\text{SC}^2}^2}{80}+\frac{1973}{630} \bigg\} \notag \\
			&\hspace{2.1cm} +\omega_6 \bigg\{ \rho_{\text{RS}^2} \left(\frac{149 \rho_{C S^2}}{300}-\frac{5 \rho_{\text{RC}^2}}{6}-\frac{91 \rho_{S^3}}{450}+\frac{5 \rho_{\text{SC}^2}}{36}+\frac{16}{45}\right)+\rho_{\text{RC}^2}	\left(\frac{\rho_{CS^2}}{4}-\frac{\rho_{S^3}}{3}+\frac{\rho_{\text{SC}^2}}{8}+\frac{1}{6}\right)+\frac{193 \rho_{C S^2}^2}{2000} \notag \\
			&\hspace{3.25cm} +\rho_{S^3} \left(-\frac{187 \rho_{CS^2}}{3000}-\frac{11 \rho_{\text{SC}^2}}{288}+\frac{91}{1800}\right)+\left(\frac{19 \rho_{\text{SC}^2}}{480}-\frac{1}{120}\right) \rho_{C S^2}+\frac{7 \rho_{\text{RS}^2}^2}{45}-\frac{581 \rho_{S^3}^2}{18000}+\frac{\rho_{\text{SC}^2}^2}{3840}-\frac{31 \rho_{\text{SC}^2}}{1440}-\frac{7}{240} \bigg\} \notag \\
			&\hspace{2.1cm} +\rho_{\text{RS}^2} \left(\frac{\rho_{C S^2}}{60}-\frac{29 \rho_{S^3}}{450}\right)+\frac{1}{150} \rho_{S^3} \rho_{C S^2}-\frac{7 \rho_{\text{RS}^2}^2}{180}-\frac{22 \rho_{S^3}^2}{1125} \Bigg] \, , \\
			t_2 &= \frac{1}{(4\pi)^3} \frac{1}{\omega_6^2} \Bigg[ \omega_6^2 \bigg\{ \rho_{\text{RS}^2}\left(\frac{7 \rho_{C S^2}}{50}-28 \rho_{\text{RC}^2}-\frac{56 \rho_{S^3}}{75}-\frac{7\rho_{\text{SC}^2}}{12}+\frac{49}{15}\right)+\rho_{\text{RC}^2} \left(\frac{21 \rho_{CS^2}}{10}+\frac{7 \rho_{S^3}}{5}+\frac{112}{5}\right)-\frac{84}{625} \rho_{C S^2}^2+\left(-\frac{7 \rho_{\text{SC}^2}}{50}-\frac{77}{125}\right) \rho_{C S^2} \notag \\
			&\hspace{3.25cm} +\rho_{S^3}\left(-\frac{154}{625} \rho_{C S^2}-\frac{77 \rho_{\text{SC}^2}}{600}-\frac{511}{750}\right)-63 \rho_{\text{RC}^2}^2-\frac{7 \rho_{\text{RS}^2}^2}{3}-\frac{1897 \rho_{S^3}^2}{7500}-\frac{7 \rho_{\text{SC}^2}^2}{320}-\frac{7 \rho_{\text{SC}^2}}{120}-\frac{48}{25} \bigg\} \notag \\
			&\hspace{2.1cm} + \omega_6 \bigg\{ \rho_{\text{RS}^2} \left(-\frac{7}{250} \rho_{C S^2}-\frac{77 \rho_{S^3}}{375}-\frac{2}{25}\right)+\frac{63 \rho_{C S^2}^2}{5000}+\rho_{S^3} \left(-\frac{77 \rho_{C S^2}}{1250}-\frac{22}{375}\right)+\frac{3 \rho_{R^3}}{2}-\frac{49 \rho_{\text{RS}^2}^2}{150}+\frac{77 \rho_{S^3}^2}{3750} \bigg\} \notag \\
			&\hspace{2.1cm} -\frac{75 \rho_{R^3}^2}{4}+\rho_{R^3} \left(-5 \rho_{\text{RS}^2}-\frac{4 \rho_{S^3}}{3}\right)+\omega_6^3 \left(\frac{42 \rho_{\text{RC}^2}}{5}+\frac{14 \rho_{\text{RS}^2}}{5}-\frac{7}{75}\right)-\frac{29 \rho_{\text{RS}^2}^2}{150}-\frac{74 \rho_{\text{RS}^2} \rho_{S^3}}{1125}-\frac{22 \rho_{S^3}^2}{16875} \Bigg] \, ,
		\end{align}
	\end{minipage}
\end{sideways}

\clearpage

\begin{sideways}
	\begin{minipage}{\textheight}
		\begin{align}
			t_3 &= \frac{1}{(4\pi)^3} \frac{1}{\omega_6^3} \Bigg[ \omega_6^3 \bigg\{ \frac{7469 \rho_{C^3}^3}{1280000}-\frac{13409 \rho_{C^3}^2}{320000}+\frac{100153 \rho_{C^3}}{2160000}-63 \rho_{\text{RC}^2}^3-\frac{7 \rho_{\text{RS}^2}^3}{3}-\frac{107 \rho_{CS^2}^3}{864000}-\frac{583 \rho_{\text{SC}^2}^3}{864000}+21 \rho_{\text{RC}^2}^2+\left(\frac{7}{5}-21 \rho_{\text{RC}^2}\right) \rho_{\text{RS}^2}^2+\frac{49 \rho_{S^3}^2}{12500} \notag \\
			&\hspace{3.25cm} +\left(\frac{673}{240000}-\frac{503 \rho_{C^3}}{192000}\right) \rho_{\text{SC}^2}^2-\frac{112 \rho_{\text{RC}^2}}{75}+\rho_{R^3} \left(-63 \rho_{\text{RC}^2}-21 \rho_{\text{RS}^2}+\frac{21}{5}\right)+\left(-63 \rho_{\text{RC}^2}^2+\frac{56 \rho_{\text{RC}^2}}{5}-\frac{14}{45}\right) \rho_{\text{RS}^2} \notag \\
			&\hspace{3.25cm} +\rho_{C S^2}^2\left(-\frac{1403 \rho_{C^3}}{960000}+\frac{101 \rho_{\text{SC}^2}}{1440000}-\frac{8591}{10800000}\right) +\rho_{S^3} \left(\frac{117 \rho_{C^3}}{4000}+\frac{279 \rho_{CS^2}}{50000}+\frac{23 \rho_{\text{SC}^2}}{2000}+\frac{857}{45000}\right) -\frac{496177}{14580000} \notag \\
			&\hspace{3.25cm} +\left(\frac{2509 \rho_{C^3}^2}{640000}-\frac{107 \rho_{C^3}}{60000}-\frac{10057}{3240000}\right) \rho_{\text{SC}^2}+\rho_{C S^2} \left(\frac{817 \rho_{C^3}^2}{640000}-\frac{9859 \rho_{C^3}}{720000}+\frac{221 \rho_{\text{SC}^2}^2}{288000}+\left(\frac{2513 \rho_{C^3}}{480000}+\frac{2809}{270000}\right) \rho_{\text{SC}^2}-\frac{5257}{1080000}\right) \bigg\} \notag \\
			& \hspace{2.1cm} + \omega_6^2 \bigg\{ \frac{\rho_{CS^2}^3}{18000}+\left(\frac{\rho_{C^3}}{20000}+\frac{7 \rho_{\text{SC}^2}}{270000}-\frac{277}{225000}\right) \rho_{C S^2}^2+\left(-\frac{47 \rho_{\text{SC}^2}^2}{648000}+\left(\frac{\rho_{C^3}}{12000}-\frac{569}{1620000}\right) \rho_{\text{SC}^2}-\frac{\rho_{C^3}}{15000}+\frac{31}{135000}\right) \rho_{CS^2} \notag \\
			&\hspace{3.25cm} -\frac{\rho_{\text{SC}^2}^3}{21600}-\frac{7 \rho_{\text{RS}^2}^2}{3375}+\frac{581 \rho_{S^3}^2}{1350000}+\left(\frac{\rho_{C^3}}{28800}+\frac{3509}{38880000}\right) \rho_{\text{SC}^2}^2+\frac{\rho_{C^3}}{45000}+\rho_{R^3}+\rho_{\text{RC}^2}^2 \left(\frac{\rho_{C^3}}{200}+\frac{\rho_{C S^2}}{180}-\frac{\rho_{\text{SC}^2}}{150}-\frac{11}{540}\right) \notag \\
			&\hspace{3.25cm} +\rho_{\text{RS}^2} \left(\frac{\rho_{\text{RC}^2}}{90}-\frac{149 \rho_{C S^2}}{22500}+\frac{91 \rho_{S^3}}{33750}-\frac{\rho_{\text{SC}^2}}{540}+\frac{134}{3375}\right)+\rho_{S^3} \left(\frac{187 \rho_{C S^2}}{225000}+\frac{11 \rho_{\text{SC}^2}}{21600}-\frac{91}{135000}\right)+\left(\frac{2423}{4860000}-\frac{\rho_{C^3}}{18000}\right) \rho_{\text{SC}^2} \notag \\
			&\hspace{3.25cm} +\rho_{\text{RC}^2} \left(\frac{1}{900} \rho_{CS^2}^2+\left(\frac{\rho_{C^3}}{1000}-\frac{11 \rho_{\text{SC}^2}}{27000}-\frac{47}{13500}\right) \rho_{C S^2}-\frac{\rho_{\text{SC}^2}^2}{900}-\frac{\rho_{C^3}}{1500}+\frac{\rho_{S^3}}{225}+\left(\frac{\rho_{C^3}}{1200}-\frac{149}{81000}\right) \rho_{\text{SC}^2}+\frac{73}{20250}\right)+\frac{77}{2430000} \bigg\} \notag \\
			&\hspace{2.1cm} + \omega_6 \bigg\{ \frac{165 \rho_{R^3}^2}{4}+\frac{11}{3} \rho_{\text{RS}^2} \rho_{R^3}+\frac{41 \rho_{\text{RS}^2}^2}{500}+\frac{22 \rho_{S^3}^2}{84375}+\rho_{\text{RS}^2}\left(\frac{29 \rho_{S^3}}{33750}-\frac{\rho_{C S^2}}{4500}\right)-\frac{\rho_{C S^2} \rho_{S^3}}{11250} \bigg\} + \frac{1125 \rho_{R^3}^3}{2}+75 \rho_{\text{RS}^2} \rho_{R^3}^2+\frac{10}{3} \rho_{\text{RS}^2}^2 \rho_{R^3}+\frac{4 \rho_{\text{RS}^2}^3}{81} \Bigg] \, ,
		\end{align}
	\end{minipage}
\end{sideways}

\clearpage

\begin{sideways}
	\begin{minipage}{\textheight}
		\begin{align}
			t_4 &= \frac{1}{(4\pi)^3} \frac{1}{\omega_6^3} \Bigg[ \omega_6^3 \bigg\{-\frac{67221 \rho_{C^3}^3}{512000}+\frac{120681 \rho_{C^3}^2}{128000}-\frac{100153 \rho_{C^3}}{96000}+\frac{107 \rho_{C S^2}^3}{38400}+\frac{583 \rho_{\text{SC}^2}^3}{38400}-7 \rho_{\text{RS}^2}^2+\frac{861 \rho_{S^3}^2}{1000}+\left(-\frac{22581 \rho_{C^3}^2}{256000}+\frac{321 \rho_{C^3}}{8000}+\frac{169657}{144000}\right) \rho_{\text{SC}^2} \notag \\
			&\hspace{3.25cm} +\rho_{C S^2}^2 \left(\frac{4209 \rho_{C^3}}{128000}-\frac{101 \rho_{\text{SC}^2}}{64000}+\frac{53731}{96000}\right)+\rho_{S^3} \left(-\frac{1053 \rho_{C^3}}{1600}+\frac{753 \rho_{C S^2}}{800}+\frac{149 \rho_{\text{SC}^2}}{160}+\frac{3697}{1200}\right)+\left(\frac{1509 \rho_{C^3}}{25600}+\frac{11281}{32000}\right) \rho_{\text{SC}^2}^2 \notag \\
			&\hspace{3.25cm} +\rho_{\text{RC}^2} \left(-\frac{1701}{400} \rho_{C S^2}^2+\left(-\frac{189 \rho_{\text{SC}^2}}{32}-\frac{63}{40}\right) \rho_{C S^2}-\frac{5229 \rho_{S^3}^2}{400}-\frac{693 \rho_{\text{SC}^2}^2}{256}+\rho_{S^3} \left(-\frac{1323}{200} \rho_{C S^2}-\frac{315 \rho_{\text{SC}^2}}{32}-\frac{273}{8}\right)-\frac{231 \rho_{\text{SC}^2}}{32}-\frac{77}{16}\right) \notag \\
			&\hspace{3.25cm} +\rho_{\text{RS}^2} \bigg[-\frac{567}{400} \rho_{C S^2}^2+\left(-\frac{63 \rho_{\text{SC}^2}}{32}-\frac{273}{200}\right) \rho_{C	S^2}-\frac{1743 \rho_{S^3}^2}{400}-\frac{231 \rho_{\text{SC}^2}^2}{256}-21 \rho_{\text{RC}^2}+\rho_{S^3} \left(-\frac{441}{200} \rho_{C S^2}-\frac{105 \rho_{\text{SC}^2}}{32}-\frac{2639}{200}\right) \notag \\
			&\hspace{3.75cm} -\frac{105 \rho_{\text{SC}^2}}{32}-\frac{1057}{240} \bigg] +\rho_{C S^2} \left(-\frac{7353 \rho_{C^3}^2}{256000}+\frac{9859 \rho_{C^3}}{32000}-\frac{221 \rho_{\text{SC}^2}^2}{12800}+\left(\frac{7901}{12000}-\frac{7539 \rho_{C^3}}{64000}\right) \rho_{\text{SC}^2}+\frac{553}{48000}\right)+\frac{572857}{648000} \bigg\} \notag \\
			&\hspace{2.1cm} + \omega_6^2 \bigg\{ \frac{7 \rho_{\text{RS}^2}^3}{4}+\left(\frac{21 \rho_{\text{RC}^2}}{4}+\frac{21 \rho_{CS^2}}{20}-\frac{7 \rho_{S^3}}{10}-\frac{77}{60}\right) \rho_{\text{RS}^2}^2-\frac{1}{800} \rho_{C S^2}^3+\frac{\rho_{\text{SC}^2}^3}{960}-\frac{889 \rho_{S^3}^2}{12000}+\left(-\frac{\rho_{C^3}}{1280}-\frac{3509}{1728000}\right) \rho_{\text{SC}^2}^2-\frac{\rho_{C^3}}{2000} \notag \\
			&\hspace{3.25cm} +\left(\frac{63}{400} \rho_{C S^2}^2-\frac{169 \rho_{C S^2}}{200}+\frac{7 \rho_{S^3}^2}{100}+\rho_{\text{RC}^2} \left(\frac{63 \rho_{C S^2}}{20}-\frac{21 \rho_{S^3}}{10}-\frac{1}{4}\right)+\left(\frac{259}{300}-\frac{21 \rho_{C S^2}}{100}\right) \rho_{S^3}+\frac{\rho_{\text{SC}^2}}{24}-\frac{7}{15}\right) \rho_{\text{RS}^2} \notag \\
			&\hspace{3.25cm} +\rho_{S^3} \left(\frac{461 \rho_{C S^2}}{2000}-\frac{11 \rho_{\text{SC}^2}}{960}-\frac{257}{1200}\right)+\rho_{C S^2}^2 \left(-\frac{9 \rho_{C^3}}{8000}-\frac{7 \rho_{\text{SC}^2}}{12000}-\frac{67}{500}\right)+\rho_{\text{RC}^2}^2\left(-\frac{9 \rho_{C^3}}{80}-\frac{\rho_{C S^2}}{8}+\frac{3 \rho_{\text{SC}^2}}{20}+\frac{11}{24}\right) \notag \\
			&\hspace{3.25cm} +\left(\frac{\rho_{C^3}}{800}-\frac{2423}{216000}\right) \rho_{\text{SC}^2}+\rho_{C S^2} \left(\frac{47 \rho_{\text{SC}^2}^2}{28800}+\left(\frac{569}{72000}-\frac{3 \rho_{C^3}}{1600}\right) \rho_{\text{SC}^2}+\frac{3 \rho_{C^3}}{2000}-\frac{31}{6000}\right)+\rho_{\text{RC}^2} \Bigg[ \frac{179}{400} \rho_{C S^2}^2+\frac{21 \rho_{S^3}^2}{100}+\frac{\rho_{\text{SC}^2}^2}{40} \notag \\
			&\hspace{3.75cm} +\frac{3 \rho_{C^3}}{200}+\left(-\frac{9 \rho_{C^3}}{400}+\frac{11 \rho_{\text{SC}^2}}{1200}+\frac{47}{600}\right) \rho_{C S^2}+\left(-\frac{63}{100} \rho_{CS^2}-\frac{1}{10}\right) \rho_{S^3}+\left(\frac{149}{3600}-\frac{3 \rho_{C^3}}{160}\right) \rho_{\text{SC}^2}-\frac{73}{900} \Bigg]-\frac{77}{108000} \bigg\} \notag \\
			&\hspace{2.1cm} + \omega_6 \bigg\{ -\frac{7 \rho_{\text{RS}^2}^3}{6}+\left(-\frac{7}{10} \rho_{CS^2}+\frac{7 \rho_{S^3}}{15}-\frac{3}{8}\right) \rho_{\text{RS}^2}^2+\left(-\frac{21}{200} \rho_{C S^2}^2+\frac{\rho_{C S^2}}{200}-\frac{7 \rho_{S^3}^2}{150}+\left(\frac{7\rho_{C S^2}}{50}-\frac{79}{900}\right) \rho_{S^3}\right) \rho_{\text{RS}^2}+\frac{28 \rho_{S^3}^2}{1125}+\frac{1}{500} \rho_{C S^2} \rho_{S^3} \notag \\
			&\hspace{3.25cm} +\rho_{R^3} \left(-\frac{105	\rho_{\text{RS}^2}^2}{4}+\left(-\frac{63}{4} \rho_{C S^2}+\frac{21 \rho_{S^3}}{2}-\frac{25}{2}\right) \rho_{\text{RS}^2}-\frac{189}{80} \rho_{C S^2}^2-\frac{21 \rho_{S^3}^2}{20}+\left(\frac{63 \rho_{C S^2}}{20}-5\right) \rho_{S^3}\right) \bigg\} \notag \\
			&\hspace{2.1cm} +\rho_{R^3} \left(5 \rho_{\text{RS}^2}^2+4 \rho_{S^3} \rho_{\text{RS}^2}+\frac{4 \rho_{S^3}^2}{5}\right) + \frac{2 \rho_{\text{RS}^2}^3}{9}+\frac{8}{45} \rho_{S^3} \rho_{\text{RS}^2}^2+\frac{8}{225} \rho_{S^3}^2 \rho_{\text{RS}^2} \Bigg] \, ,
		\end{align}
	\end{minipage}
\end{sideways}

\clearpage

\begin{sideways}
	\begin{minipage}{\textheight}
		\begin{align}
			t_5 &= \frac{1}{(4\pi)^3} \frac{1}{\omega_6^2} \Bigg[ \omega_6^2 \bigg\{\frac{22407 \rho_{C^3}^3}{512000}+\frac{227133 \rho_{C^3}^2}{128000}-\frac{1040207 \rho_{C^3}}{288000}-\frac{107 \rho_{C S^2}^3}{115200}-\frac{583 \rho_{\text{SC}^2}^3}{115200}+\left(\frac{2713}{32000}-\frac{503 \rho_{C^3}}{25600}\right) \rho_{\text{SC}^2}^2 +\left(\frac{7527 \rho_{C^3}^2}{256000}+\frac{8631 \rho_{C^3}}{16000}+\frac{50603}{432000}\right) \rho_{\text{SC}^2} \notag \\
			&\hspace{3.25cm} -21 \rho_{\text{RC}^2}^2-\frac{343 \rho_{S^3}^2}{5000}+\rho_{S^3} \left(-\frac{819 \rho_{C^3}}{1600}-\frac{1953 \rho_{C S^2}}{20000}-\frac{161 \rho_{\text{SC}^2}}{800}-\frac{5999}{18000}\right)+\rho_{C S^2}^2 \left(-\frac{1403 \rho_{C^3}}{128000}+\frac{101 \rho_{\text{SC}^2}}{192000}+\frac{18529}{1440000}\right) \notag \\
			&\hspace{3.25cm} +\rho_{\text{RC}^2} \left(-\frac{15039 \rho_{C^3}^2}{1280}+\frac{3357 \rho_{C^3}}{160}-\frac{159}{320} \rho_{C S^2}^2-\frac{75 \rho_{\text{SC}^2}^2}{64}+\rho_{C S^2}	\left(-\frac{801 \rho_{C^3}}{320}+\frac{21 \rho_{\text{SC}^2}}{32}-\frac{17}{16}\right)+\left(-\frac{261 \rho_{C^3}}{64}-\frac{53}{16}\right) \rho_{\text{SC}^2}+\frac{8467}{240}\right) \notag \\
			&\hspace{3.25cm} +\rho_{\text{RS}^2} \left(-\frac{5013 \rho_{C^3}^2}{1280}+\frac{1147 \rho_{C^3}}{160}-\frac{53}{320} \rho_{C S^2}^2-\frac{25 \rho_{\text{SC}^2}^2}{64}-7 \rho_{\text{RC}^2}+\rho_{C S^2} \left(-\frac{267 \rho_{C^3}}{320}+\frac{7 \rho_{\text{SC}^2}}{32}-\frac{43}{80}\right)+\left(-\frac{87 \rho_{C^3}}{64}-\frac{37}{48}\right) \rho_{\text{SC}^2}+\frac{581}{48}\right) \notag \\
			&\hspace{3.25cm} +\rho_{C S^2} \left(\frac{2451 \rho_{C^3}^2}{256000}+\frac{38261 \rho_{C^3}}{96000}+\frac{221 \rho_{\text{SC}^2}^2}{38400}+\left(\frac{2513 \rho_{C^3}}{64000}-\frac{13027}{72000}\right) \rho_{\text{SC}^2}+\frac{5381}{48000}\right)-\frac{53224519}{13608000} \bigg\} \notag \\
			&\hspace{2.1cm} + \omega_6 \bigg\{\frac{15 \rho_{\text{RC}^2}^3}{4}+\left(\frac{3 \rho_{C^3}}{80}+\frac{19 \rho_{C S^2}}{24}+\frac{23 \rho	_{\text{SC}^2}}{40}-\frac{149}{72}\right) \rho_{\text{RC}^2}^2+\frac{\rho_{C S^2}^3}{2400}-\frac{\rho_{\text{SC}^2}^3}{2880}+\frac{49 \rho_{\text{RS}^2}^2}{1350}-\frac{4067 \rho_{S^3}^2}{540000}+\left(\frac{\rho_{C^3}}{3840}-\frac{49921}{5184000}\right) \rho_{\text{SC}^2}^2 \notag \\
			&\hspace{3.25cm} +\left(\frac{11}{240} \rho_{C S^2}^2+\left(\frac{3 \rho_{C^3}}{400}+\frac{107 \rho	_{\text{SC}^2}}{1800}-\frac{557}{1800}\right) \rho_{C S^2}+\frac{17 \rho_{\text{SC}^2}^2}{960}-\frac{\rho_{C^3}}{200}-\frac{7 \rho_{S^3}}{90}+\left(\frac{\rho_{C^3}}{160}-\frac{2879}{10800}\right) \rho_{\text{SC}^2}-\frac{68}{675}\right) \rho_{\text{RC}^2} \notag \\
			&\hspace{3.25cm}+\frac{\rho_{C^3}}{6000}+\frac{\rho_{R^3}}{12} +\rho_{S^3} \left(-\frac{1309 \rho_{C S^2}}{90000}-\frac{77 \rho_{\text{SC}^2}}{8640}+\frac{637}{54000}\right)+\rho_{C S^2}^2 \left(\frac{3 \rho_{C^3}}{8000}+\frac{7 \rho_{\text{SC}^2}}{36000}+\frac{263}{30000}\right)+\left(\frac{8933}{648000}-\frac{\rho_{C^3}}{2400}\right) \rho_{\text{SC}^2} \notag \\
			&\hspace{3.25cm} +\rho_{\text{RS}^2} \left(\frac{5 \rho_{\text{RC}^2}^2}{4}+\left(\frac{\rho_{C S^2}}{4}+\frac{5 \rho_{\text{SC}^2}}{24}-\frac{13}{36}\right) \rho_{\text{RC}^2}+\frac{1}{80} \rho_{C S^2}^2+\frac{5 \rho_{\text{SC}^2}^2}{576}-\frac{637 \rho_{S^3}}{13500}+\rho_{C S^2} \left(\frac{\rho_{\text{SC}^2}}{48}+\frac{893}{9000}\right)+\frac{\rho_{\text{SC}^2}}{54}+\frac{83}{900}\right) \notag \\
			&\hspace{3.25cm} +\rho_{C S^2}\left(-\frac{47 \rho_{\text{SC}^2}^2}{86400}+\left(\frac{\rho_{C^3}}{1600}-\frac{3059}{216000}\right)\rho_{\text{SC}^2}-\frac{\rho_{C^3}}{2000}+\frac{841}{18000}\right)+\frac{16967}{324000} \bigg\} \notag \\
			&\hspace{2.1cm} -\frac{49 \rho_{\text{RS}^2}^2}{5400}+\left(-\frac{5 \rho_{\text{RC}^2}^2}{6}+\left(-\frac{1}{6} \rho_{C S^2}-\frac{5 \rho_{\text{SC}^2}}{36}-\frac{2}{9}\right) \rho_{\text{RC}^2}-\frac{1}{120} \rho_{C S^2}^2-\frac{5 \rho_{\text{SC}^2}^2}{864}-\frac{203 \rho_{S^3}}{13500}+\rho_{C S^2} \left(\frac{67}{1800}-\frac{\rho_{\text{SC}^2}}{72}\right)+\frac{\rho_{\text{SC}^2}}{108}+\frac{1}{18}\right) \rho_{\text{RS}^2} \notag \\
			&\hspace{2.1cm} -\frac{77 \rho_{S^3}^2}{16875} +\frac{7 \rho_{C S^2} \rho_{S^3}}{4500}+\rho_{R^3} \left(-\frac{75 \rho_{\text{RC}^2}^2}{4}+\left(-\frac{15}{4} \rho_{C S^2}-\frac{25 \rho_{\text{SC}^2}}{8}-5\right)	\rho_{\text{RC}^2}-\frac{3}{16} \rho_{C S^2}^2-\frac{25 \rho_{\text{SC}^2}^2}{192}+\rho_{C S^2} \left(\frac{3}{4}-\frac{5 \rho_{\text{SC}^2}}{16}\right)+\frac{5 \rho_{\text{SC}^2}}{24}+\frac{5}{4}\right) \Bigg] \, ,
		\end{align}
	\end{minipage}
\end{sideways}

\clearpage

\begin{sideways}
	\begin{minipage}{\textheight}
		\begin{align}
			t_6 &= \frac{1}{(4\pi)^3} \frac{1}{\omega_6^3} \Bigg[ \omega_6^3 \bigg\{ \frac{67221\rho_{C^3}^3}{204800}-\frac{120681 \rho_{C^3}^2}{51200}+\frac{100153 \rho_{C^3}}{38400}-\frac{380791 \rho_{C S^2}^3}{1920000}+\left(\frac{21411}{51200}-\frac{1509 \rho_{C^3}}{10240}\right) \rho_{\text{SC}^2}^2+\left(\frac{22581 \rho_{C^3}^2}{102400}-\frac{321 \rho_{C^3}}{3200}+\frac{95491}{115200}\right) \rho_{\text{SC}^2} \notag \\
			&\hspace{3.25cm} -\frac{126903 \rho_{S^3}^3}{80000}-\frac{11783	\rho_{\text{SC}^2}^3}{122880} +\rho_{S^3}^2 \left(-\frac{65583 \rho_{C	S^2}}{80000}-\frac{819 \rho_{\text{SC}^2}}{512}-\frac{108213}{16000}\right)+\rho_{C S^2}^2 \left(-\frac{4209 \rho_{C^3}}{51200}-\frac{5569 \rho_{\text{SC}^2}}{25600}+\frac{533461}{192000}\right) \notag \\
			&\hspace{3.25cm} +\rho_{S^3} \left(\frac{20979 \rho_{C S^2}^2}{80000}+\left(\frac{243}{1000}-\frac{189 \rho_{\text{SC}^2}}{1280}\right) \rho_{C S^2}-\frac{4473 \rho_{\text{SC}^2}^2}{10240}+\frac{1053 \rho_{C^3}}{640}-\frac{3687 \rho_{\text{SC}^2}}{1280}-\frac{9137}{3200}\right) \notag \\
			&\hspace{3.25cm} +\rho_{C S^2} \left(\frac{7353 \rho_{C^3}^2}{102400}-\frac{9859 \rho_{C^3}}{12800}-\frac{881 \rho_{\text{SC}^2}^2}{10240}+\left(\frac{7539 \rho_{C^3}}{25600}+\frac{65101}{19200}\right) \rho_{\text{SC}^2}+\frac{70973}{19200}\right)+\frac{24287}{64800} \bigg\} \notag \\
			&\hspace{2.1cm} + \omega_6^2 \bigg\{ \frac{2047 \rho_{C S^2}^3}{20000}+\left(\frac{9 \rho_{C^3}}{3200}+\frac{2863 \rho_{\text{SC}^2}}{19200}-\frac{1751}{8000}\right) \rho_{C S^2}^2+\left(-\frac{47 \rho_{\text{SC}^2}^2}{11520}+\left(\frac{3 \rho_{C^3}}{640}-\frac{569}{28800}\right) \rho_{\text{SC}^2}-\frac{3 \rho_{C^3}}{800}+\frac{31}{2400}\right) \rho_{C S^2} \notag \\
			&\hspace{3.25cm} +\frac{1743 \rho_{S^3}^3}{10000}-\frac{\rho_{\text{SC}^2}^3}{384}+\left(\frac{\rho_{C^3}}{512}+\frac{3509}{691200}\right) \rho_{\text{SC}^2}^2+\frac{\rho_{C^3}}{800}+\rho_{\text{RC}^2}^2 \left(\frac{9 \rho_{C^3}}{32}+\frac{5 \rho_{C S^2}}{16}-\frac{3 \rho_{\text{SC}^2}}{8}-\frac{55}{48}\right) \notag \\
			&\hspace{3.25cm} +\rho_{\text{RS}^2} \left(\frac{1323 \rho_{C S^2}^2}{2000}+\left(\frac{63 \rho_{\text{SC}^2}}{64}+\frac{167}{200}\right) \rho_{C	S^2}-\frac{1743 \rho_{S^3}^2}{1000}+\frac{5 \rho_{\text{RC}^2}}{8}+\rho_{S^3} \left(\frac{4347 \rho_{C S^2}}{2000}-\frac{21 \rho_{\text{SC}^2}}{32}+\frac{511}{150}\right)-\frac{5 \rho_{\text{SC}^2}}{48}-\frac{4}{15}\right) \notag \\
			&\hspace{3.25cm} +\rho_{S^3} \left(\frac{2079 \rho_{CS^2}^2}{8000}+\left(\frac{4513}{4000}-\frac{63 \rho_{\text{SC}^2}}{320}\right) \rho_{C S^2}+\frac{11 \rho_{\text{SC}^2}}{384}-\frac{91}{2400}\right)+\rho_{S^3}^2\left(-\frac{1197 \rho_{C S^2}}{2500}+\frac{21 \rho_{\text{SC}^2}}{320}-\frac{9751}{24000}\right) \notag \\
			&\hspace{3.25cm} +\rho_{\text{RC}^2} \left(\frac{1}{16} \rho_{C S^2}^2+\left(\frac{9 \rho_{C^3}}{160}-\frac{11 \rho_{\text{SC}^2}}{480}-\frac{47}{240}\right) \rho_{C S^2}-\frac{\rho_{\text{SC}^2}^2}{16}-\frac{3 \rho_{C^3}}{80}+\frac{\rho_{S^3}}{4}+\left(\frac{3 \rho_{C^3}}{64}-\frac{149}{1440}\right) \rho_{\text{SC}^2}+\frac{73}{360}\right) \notag \\
			&\hspace{3.25cm} +\left(\frac{2423}{86400}-\frac{\rho_{C^3}}{320}\right) \rho_{\text{SC}^2}+\rho_{\text{RS}^2}^2 \left(\frac{441 \rho_{C S^2}}{400}+\frac{1743 \rho_{S^3}}{400}+\frac{105 \rho_{\text{SC}^2}}{64}+\frac{1673}{240}\right) +\frac{77}{43200} \bigg\} \notag \\
			&\hspace{2.1cm} + \omega_6 \bigg\{ \frac{7 \rho_{\text{RS}^2}^3}{10}+\left(\frac{21 \rho_{C S^2}}{50}-\frac{9}{80}\right) \rho_{\text{RS}^2}^2+\left(\frac{63 \rho_{C S^2}^2}{1000}-\frac{\rho_{C S^2}}{80}-\frac{21 \rho_{S^3}^2}{250}+\left(\frac{21 \rho_{C S^2}}{250}-\frac{13}{200}\right) \rho_{S^3}\right) \rho_{\text{RS}^2}+\frac{7 \rho_{S^3}^3}{625} \notag \\
			&\hspace{3.25cm} +\left(-\frac{21}{625} \rho_{C S^2}-\frac{1}{125}\right) \rho_{S^3}^2+\left(\frac{63 \rho_{CS^2}^2}{2500}-\frac{\rho_{C S^2}}{200}\right) \rho_{S^3} \bigg\} + \frac{8 \rho_{\text{RS}^2}^3}{45}+\frac{16}{75} \rho_{S^3} \rho_{\text{RS}^2}^2+\frac{32}{375} \rho_{S^3}^2 \rho_{\text{RS}^2}+\frac{64 \rho_{S^3}^3}{5625} \Bigg] \, ,
		\end{align}
	\end{minipage}
\end{sideways}

\clearpage

\begin{sideways}
	\begin{minipage}{\textheight}
		\begin{align}
			t_7 &= \frac{1}{(4\pi)^3} \frac{1}{\omega_6^2} \Bigg[ \omega_6^2 \bigg\{ \frac{67221 \rho_{C^3}^3}{102400}-\frac{120681 \rho_{C^3}^2}{25600}+\frac{19357 \rho_{C^3}}{4800}+\frac{221 \rho_{C S^2}^3}{38400}+\left(-\frac{19671 \rho_{C^3}}{20480}-\frac{23999}{25600}\right) \rho_{\text{SC}^2}^2+\left(\frac{22581 \rho_{C^3}^2}{51200}-\frac{25293 \rho_{C^3}}{12800}-\frac{48907}{28800}\right) \rho_{\text{SC}^2} \notag \\
			&\hspace{3.25cm} -\frac{13807 \rho_{\text{SC}^2}^3}{30720}+\rho_{S^3}^2 \left(\frac{9153 \rho_{C^3}}{6400}+\frac{2727 \rho_{C S^2}}{3200}-\frac{513 \rho_{\text{SC}^2}}{640}+\frac{137}{1600}\right)+\rho_{C S^2}^2 \left(-\frac{37581 \rho_{C^3}}{25600}-\frac{4489 \rho_{\text{SC}^2}}{12800}-\frac{23659}{19200}\right) \notag \\
			&\hspace{3.25cm} +\rho_{S^3} \left(\frac{549 \rho_{C S^2}^2}{1600}+\left(-\frac{10989 \rho_{C^3}}{3200}-\frac{1359 \rho_{\text{SC}^2}}{1280}+\frac{3383}{1600}\right) \rho_{C S^2}-\frac{315 \rho_{\text{SC}^2}^2}{256}+\frac{6531 \rho_{C^3}}{640}+\left(\frac{73}{128}-\frac{513 \rho_{C^3}}{512}\right) \rho_{\text{SC}^2}+\frac{121}{480}\right) \notag \\
			&\hspace{3.25cm} +\rho_{C S^2}\left(\frac{7353 \rho_{C^3}^2}{51200}-\frac{54709 \rho_{C^3}}{6400}-\frac{7081 \rho_{\text{SC}^2}^2}{10240}+\left(-\frac{2517 \rho_{C^3}}{1600}-\frac{23383}{19200}\right) \rho			_{\text{SC}^2}-\frac{317}{400}\right)-\frac{9577}{8100} \bigg\} \notag \\
			&\hspace{2.1cm} + \omega_6 \bigg\{ \frac{277 \rho_{C S^2}^3}{1600}+\left(\frac{819 \rho_{C^3}}{3200}+\frac{97 \rho_{\text{SC}^2}}{2400}+\frac{313}{800}\right) \rho_{C S^2}^2+\left(\frac{89 \rho_{\text{SC}^2}^2}{2880}+\left(\frac{3 \rho_{C^3}}{320}+\frac{931}{14400}\right) \rho_{\text{SC}^2}-\frac{3 \rho_{C^3}}{400}-\frac{53}{400}\right)\rho_{C S^2}-\frac{\rho_{\text{SC}^2}^3}{192} \notag \\
			&\hspace{3.25cm} +\left(\frac{\rho_{C^3}}{256}+\frac{3509}{345600}\right) \rho_{\text{SC}^2}^2+\frac{\rho_{C^3}}{400}+\rho_{\text{RC}^2}^2\left(\frac{9 \rho_{C^3}}{16}+\frac{5 \rho_{C S^2}}{8}-\frac{3 \rho_{\text{SC}^2}}{4}-\frac{55}{24}\right)+\rho_{\text{RS}^2}^2 \left(\frac{89 \rho_{C^3}}{32}+\frac{53 \rho_{C S^2}}{48}-\frac{35 \rho_{\text{SC}^2}}{48}+\frac{145}{72}\right) \notag \\
			&\hspace{3.25cm} +\rho_{S^3}^2 \left(\frac{89 \rho_{C^3}}{800}+\frac{11 \rho_{C S^2}}{1200}-\frac{7 \rho_{\text{SC}^2}}{120}+\frac{1567}{7200}\right)+\left(\frac{2423}{43200}-\frac{\rho_{C^3}}{160}\right) \rho_{\text{SC}^2} +\rho_{\text{RC}^2} \bigg[ \frac{4}{5} \rho_{CS^2}^2+\left(\frac{9 \rho_{C^3}}{80}+\frac{203 \rho_{\text{SC}^2}}{480}+\frac{37}{30}\right) \rho_{C S^2} \notag \\
			&\hspace{3.75cm} -\frac{7 \rho_{S^3}^2}{20}-\frac{\rho_{\text{SC}^2}^2}{8}-\frac{3\rho_{C^3}}{40}+\rho_{S^3} \left(\frac{3 \rho_{C S^2}}{40}-\frac{5 \rho_{\text{SC}^2}}{16}-\frac{19}{12}\right)+\left(\frac{3 \rho_{C^3}}{32}-\frac{149}{720}\right) \rho_{\text{SC}^2}+\frac{73}{180} \bigg] \notag \\
			&\hspace{3.25cm} +\rho_{S^3} \left(-\frac{1}{8} \rho_{C S^2}^2+\left(-\frac{267 \rho_{C^3}}{800}+\frac{\rho_{\text{SC}^2}}{16}-\frac{67}{80}\right) \rho_{CS^2}-\frac{5 \rho_{\text{SC}^2}^2}{192}-\frac{55 \rho_{\text{SC}^2}}{576}+\frac{37}{240}\right) +\rho_{\text{RS}^2} \bigg[\frac{71}{80} \rho_{C S^2}^2+\left(\frac{267 \rho_{C^3}}{160}-\frac{3 \rho_{\text{SC}^2}}{32}+\frac{69}{40}\right) \rho_{C S^2} \notag \\
			&\hspace{3.75cm} +\frac{25 \rho_{\text{SC}^2}^2}{192}+\rho_{S^3} \left(-\frac{89 \rho_{C^3}}{80}-\frac{4 \rho_{CS^2}}{15}+\frac{7 \rho_{\text{SC}^2}}{16}-\frac{481}{180}\right)-\frac{5 \rho_{\text{SC}^2}}{72}+\rho_{\text{RC}^2} \left(\frac{9 \rho_{C S^2}}{4}+\frac{7 \rho_{S^3}}{4}+\frac{25 \rho_{\text{SC}^2}}{16}+\frac{25}{6}\right)-\frac{3}{4}\bigg]+\frac{77}{21600} \bigg\} \notag \\
			&\hspace{2.1cm} +\left(-\frac{1}{10} \rho_{CS^2}^2+\left(\frac{11}{120}-\frac{\rho_{\text{SC}^2}}{12}\right) \rho_{C S^2}+\rho_{\text{RC}^2} \left(-\rho_{C S^2}-\frac{2 \rho_{S^3}}{3}\right)+\rho_{S^3}\left(-\frac{1}{15} \rho_{C S^2}-\frac{\rho_{\text{SC}^2}}{18}+\frac{7}{60}\right)\right) \rho_{\text{RS}^2}+\rho_{S^3}^2 \left(\frac{2 \rho_{C S^2}}{75}+\frac{\rho_{\text{SC}^2}}{45}-\frac{1}{75}\right) \notag \\
			&\hspace{2.1cm} +\left(-\frac{10 \rho_{\text{RC}^2}}{3}-\frac{\rho_{C S^2}}{3}-\frac{5 \rho_{\text{SC}^2}}{18}+\frac{3}{8}\right) \rho_{\text{RS}^2}^2 +\rho_{\text{RC}^2} \left(\frac{4 \rho_{S^3}^2}{15}-\frac{2}{5} \rho_{C S^2} \rho_{S^3}\right)+\rho_{S^3} \left(\rho_{C S^2} \left(\frac{11}{300}-\frac{\rho_{\text{SC}^2}}{30}\right)-\frac{1}{25} \rho_{C S^2}^2\right) \Bigg] \, ,
		\end{align}
	\end{minipage}
\end{sideways}

\clearpage

\begin{sideways}
	\begin{minipage}{\textheight}
		\begin{align}
			t_8 &= \frac{1}{(4\pi)^3} \frac{1}{\omega_6^2} \Bigg[ \omega_6^2 \bigg\{-\frac{67221 \rho_{C^3}^3}{51200}+\frac{837 \rho_{C^3}^2}{200}-\frac{88663 \rho_{C^3}}{9600} +\rho_{C S^2}^2\left(-\frac{35871 \rho_{C^3}}{64000}-\frac{4141 \rho_{\text{SC}^2}}{12800}-\frac{22549}{12000}\right)+\left(-\frac{246501 \rho_{C^3}^2}{51200}+\frac{243 \rho_{C^3}}{6400}+\frac{177419}{28800}\right) \rho_{\text{SC}^2} \notag \\
			&\hspace{3.25cm} +\left(-\frac{57 \rho_{C^3}}{160}-\frac{649}{640}\right) \rho_{\text{SC}^2}^2 +\rho_{C S^2} \left(-\frac{641187 \rho_{C^3}^2}{128000}+\frac{8563 \rho_{C^3}}{8000}-\frac{47 \rho_{\text{SC}^2}^2}{128}+\left(-\frac{52599 \rho_{C^3}}{12800}-\frac{39821}{9600}\right) \rho_{\text{SC}^2}+\frac{40187}{8000}\right) \notag \\
			&\hspace{3.25cm} +\rho_{S^3} \Bigg[-\frac{402273 \rho_{C^3}^2}{64000}-\frac{15591 \rho_{C^3}}{8000}-\frac{6033 \rho_{C S^2}^2}{16000}-\frac{129 \rho_{\text{SC}^2}^2}{640}+\rho_{C S^2} \left(-\frac{43047 \rho_{C^3}}{16000}+\frac{15 \rho_{\text{SC}^2}}{64}-\frac{13449}{4000}\right) \notag \\
			&\hspace{3.75cm} +\left(-\frac{99 \rho_{C^3}}{640}-\frac{2433}{800}\right) \rho_{\text{SC}^2}+\frac{80329}{12000} \Bigg] +\frac{37 \rho_{\text{SC}^2}^3}{1536}-\frac{511 \rho_{CS^2}^3}{96000}-\frac{147 \rho_{S^3}^2}{100} -\frac{368113}{56700} \bigg\} \notag \\
			&\hspace{2.1cm} + \omega_6 \bigg\{-\frac{43 \rho_{C S^2}^3}{2000}+\left(\frac{369 \rho_{C^3}}{2000}+\frac{133 \rho_{\text{SC}^2}}{960}+\frac{1079}{2000}\right)\rho_{C S^2}^2 +\left(\frac{461 \rho_{\text{SC}^2}^2}{2880}+\left(\frac{231 \rho_{C^3}}{1600}+\frac{1313}{7200}\right) \rho_{\text{SC}^2}-\frac{231 \rho_{C^3}}{2000}-\frac{1247}{3000}\right) \rho_{C S^2}+\frac{19 \rho_{\text{SC}^2}^3}{768} \notag \\
			&\hspace{3.25cm} +\frac{7 \rho_{\text{RS}^2}^2}{9}-\frac{581 \rho_{S^3}^2}{3600}+\left(-\frac{\rho_{C^3}}{128}-\frac{931}{34560}\right) \rho_{\text{SC}^2}^2-\frac{\rho_{C^3}}{200}+\left(\frac{\rho_{C^3}}{80}-\frac{3983}{21600}\right) \rho_{\text{SC}^2}+\rho_{\text{RC}^2}^2 \left(-\frac{9 \rho_{C^3}}{8}+\rho_{C S^2}+\frac{15 \rho_{S^3}}{4}+\frac{57 \rho_{\text{SC}^2}}{16}+\frac{22}{3}\right) \notag \\
			&\hspace{3.25cm} +\rho_{S^3} \left(\frac{117 \rho_{CS^2}^2}{2000}+\left(-\frac{261 \rho_{C^3}}{2000}+\frac{\rho_{\text{SC}^2}}{200}-\frac{1229}{3000}\right) \rho_{C S^2}-\frac{7 \rho_{\text{SC}^2}^2}{192}+\frac{87 \rho_{C^3}}{1000}+\left(-\frac{87 \rho_{C^3}}{800}-\frac{371}{1440}\right) \rho_{\text{SC}^2}+\frac{461}{1000}\right) \notag \\
			&\hspace{3.25cm} +\rho_{\text{RC}^2} \Bigg[-\frac{23}{200} \rho_{CS^2}^2+\left(\frac{693 \rho_{C^3}}{400}+\frac{481 \rho_{\text{SC}^2}}{240}+\frac{1183}{300}\right) \rho_{C S^2}+\frac{19 \rho_{\text{SC}^2}^2}{32}+\frac{3 \rho_{C^3}}{20}+\rho_{S^3} \left(-\frac{261 \rho_{C^3}}{200}+\frac{24 \rho_{C S^2}}{25}-\frac{\rho_{\text{SC}^2}}{8}-\frac{293}{75}\right) \notag \\
			&\hspace{3.75cm} +\left(\frac{269}{360}-\frac{3 \rho_{C^3}}{16}\right) \rho_{\text{SC}^2}-\frac{44}{45} \Bigg] +\rho_{\text{RS}^2} \Bigg[-\frac{21}{200} \rho_{C S^2}^2+\left(\frac{261 \rho_{C^3}}{400}+\frac{23 \rho_{\text{SC}^2}}{80}+\frac{817}{300}\right) \rho_{C S^2}+\frac{5 \rho_{\text{SC}^2}^2}{16}-\frac{87 \rho_{C^3}}{200}-\frac{91 \rho_{S^3}}{90} \notag \\
			&\hspace{3.75cm} +\left(\frac{87 \rho_{C^3}}{160}+\frac{59}{72}\right) \rho_{\text{SC}^2}+\rho_{\text{RC}^2} \left(\frac{261 \rho_{C^3}}{40}-\frac{21 \rho_{C S^2}}{20}+\frac{15 \rho_{\text{SC}^2}}{4}+\frac{68}{15}\right)+\frac{103}{150} \Bigg]-\frac{623}{10800} \bigg\} \notag \\
			&\hspace{2.1cm} -\frac{4}{5} \rho_{S^3} \rho_{\text{RC}^2}^2+\rho_{S^3} \left(-\frac{4}{25} \rho_{C S^2}-\frac{2 \rho_{\text{SC}^2}}{15}+\frac{8}{75}\right) \rho_{\text{RC}^2}-\frac{7 \rho_{\text{RS}^2}^2}{36}-\frac{22 \rho_{S^3}^2}{225}+\rho_{S^3} \left(-\frac{1}{125} \rho_{CS^2}^2+\left(\frac{43}{750}-\frac{\rho_{\text{SC}^2}}{75}\right) \rho_{C S^2}-\frac{\rho_{\text{SC}^2}^2}{180}+\frac{7 \rho_{\text{SC}^2}}{450}+\frac{2}{375}\right) \notag  \\
			&\hspace{2.1cm} +\rho_{\text{RS}^2} \left(-2 \rho_{\text{RC}^2}^2+\left(-\frac{2}{5} \rho_{C S^2}-\frac{\rho_{\text{SC}^2}}{3}+\frac{4}{15}\right) \rho_{\text{RC}^2}-\frac{1}{50} \rho_{C S^2}^2-\frac{\rho_{\text{SC}^2}^2}{72}-\frac{29 \rho_{S^3}}{90}+\rho_{C S^2}\left(\frac{43}{300}-\frac{\rho_{\text{SC}^2}}{30}\right)+\frac{7 \rho_{\text{SC}^2}}{180}+\frac{1}{75}\right) \Bigg] \, ,
		\end{align}
	\end{minipage}
\end{sideways}

\clearpage

\begin{sideways}
	\begin{minipage}{\textheight}
		\begin{align}
			t_9 &= \frac{1}{(4\pi)^3} \frac{1}{\omega_6^2} \Bigg[ \omega_6^2 \bigg\{-\frac{14157 \rho_{C^3}^3}{25600}-\frac{92763 \rho_{C^3}^2}{6400}+\frac{33649 \rho_{C^3}}{4800}+\rho_{C S^2}^2			\left(-\frac{87 \rho_{C^3}}{6400}+\frac{203 \rho_{\text{SC}^2}}{3200}-\frac{2033}{24000}\right)+\rho_{S^3} \left(\frac{1053 \rho_{C^3}}{160}+\frac{2511 \rho_{CS^2}}{2000}+\frac{207 \rho_{\text{SC}^2}}{80}+\frac{857}{200}\right) \notag \\
			&\hspace{3.25cm} +\rho_{C S^2} \left(\frac{3519 \rho_{C^3}^2}{12800}-\frac{7327 \rho_{C^3}}{1600}+\left(\frac{717 \rho_{C^3}}{3200}+\frac{7481}{4800}\right) \rho_{\text{SC}^2}-\frac{37 \rho_{\text{SC}^2}^2}{640}-\frac{5221}{2400}\right)+\left(\frac{1647}{1600}-\frac{627 \rho_{C^3}}{1280}\right) \rho_{\text{SC}^2}^2 \notag \\
			&\hspace{3.25cm} +\left(-\frac{12357 \rho_{C^3}^2}{12800}-\frac{4899 \rho_{C^3}}{3200}-\frac{197}{900}\right) \rho_{\text{SC}^2}-\frac{29 \rho_{CS^2}^3}{1920}+\frac{441 \rho_{S^3}^2}{500}-\frac{49 \rho_{\text{SC}^2}^3}{1920}+\frac{1265449}{113400} \bigg\} \notag \\
			&\hspace{2.1cm} +\omega_6 \bigg\{ \rho_{\text{RC}^2}^2 \left(\frac{63 \rho_{C^3}}{16}+\frac{5 \rho_{C S^2}}{8}+\frac{3 \rho_{\text{SC}^2}}{8}-\frac{25}{6}\right)+\rho_{C S^2} \left(-\frac{21 \rho_{C^3}}{400}+\left(\frac{21 \rho_{C^3}}{320}-\frac{649}{7200}\right) \rho_{\text{SC}^2}+\frac{61 \rho_{\text{SC}^2}^2}{5760}+\frac{13}{300}\right) \notag \\
			&\hspace{3.25cm} +\rho_{\text{RC}^2} \left(-\frac{21 \rho_{C^3}}{40}+\rho_{C S^2} \left(\frac{63 \rho_{C^3}}{80}+\frac{43 \rho_{\text{SC}^2}}{240}-\frac{37}{60}\right)+\left(\frac{21 \rho_{C^3}}{32}-\frac{107}{180}\right) \rho_{\text{SC}^2}+\frac{1}{8} \rho_{C S^2}^2+\rho_{S^3}+\frac{\rho_{\text{SC}^2}^2}{16}+\frac{34}{45}\right) \notag \\
			&\hspace{3.25cm} +\frac{7 \rho_{C^3}}{400}+\left(\frac{7 \rho_{C^3}}{256}+\frac{409}{172800}\right) \rho_{\text{SC}^2}^2+\left(\frac{2503}{21600}-\frac{7 \rho_{C^3}}{160}\right) \rho_{\text{SC}^2}+\rho_{\text{RS}^2} \left(-\frac{149 \rho_{C S^2}}{100}+\frac{5 \rho_{\text{RC}^2}}{2}+\frac{91 \rho_{S^3}}{150}-\frac{5 \rho_{\text{SC}^2}}{12}-\frac{16}{15}\right) \notag \\
			&\hspace{3.25cm} +\rho_{C S^2}^2 \left(\frac{63	\rho_{C^3}}{1600}+\frac{17 \rho_{\text{SC}^2}}{1200}-\frac{529}{2000}\right)+\frac{1}{160} \rho_{C S^2}^3+\rho_{S^3} \left(\frac{187 \rho_{C	S^2}}{1000}+\frac{11 \rho_{\text{SC}^2}}{96}-\frac{91}{600}\right)-\frac{7 \rho_{\text{RS}^2}^2}{15}+\frac{581 \rho_{S^3}^2}{6000}+\frac{\rho_{\text{SC}^2}^3}{384}+\frac{97}{10800} \bigg\} \notag \\
			&\hspace{2.1cm} +\rho_{\text{RS}^2} \left(\frac{29 \rho_{S^3}}{150}-\frac{\rho_{CS^2}}{20}\right)-\frac{1}{50} \rho_{S^3} \rho_{C S^2}+\frac{7 \rho_{\text{RS}^2}^2}{60}+\frac{22 \rho_{S^3}^2}{375} \Bigg] \, , \\
			t_{10} &= \frac{1}{(4\pi)^3} \frac{1}{\omega_6^2} \Bigg[ \omega_6^2 \bigg\{-\frac{22407 \rho_{C^3}^3}{102400}+\frac{40227 \rho_{C^3}^2}{25600}-\frac{100153 \rho_{C^3}}{57600}+\rho_{S^3} \left(-\frac{351 \rho_{C^3}}{320}-\frac{837 \rho_{C	S^2}}{4000}-\frac{69 \rho_{\text{SC}^2}}{160}-\frac{857}{1200}\right)+\rho_{C S^2}^2 \left(\frac{1403 \rho_{C^3}}{25600}-\frac{101 \rho_{\text{SC}^2}}{38400}+\frac{8591}{288000}\right) \notag \\
			&\hspace{3.25cm} +\rho_{C S^2} \left(-\frac{2451 \rho_{C^3}^2}{51200}+\frac{9859 \rho_{C^3}}{19200}+\left(-\frac{2513 \rho_{C^3}}{12800}-\frac{2809}{7200}\right) \rho_{\text{SC}^2}-\frac{221 \rho_{\text{SC}^2}^2}{7680}+\frac{5257}{28800}\right)+\left(\frac{503 \rho_{C^3}}{5120}-\frac{673}{6400}\right) \rho_{\text{SC}^2}^2 \notag \\
			&\hspace{3.25cm} +\left(-\frac{7527 \rho_{C^3}^2}{51200}+\frac{107 \rho_{C^3}}{1600}+\frac{10057}{86400}\right) \rho_{\text{SC}^2}+\frac{107 \rho_{C S^2}^3}{23040}-\frac{147 \rho_{S^3}^2}{1000}+\frac{583 \rho_{\text{SC}^2}^3}{23040}-\frac{5831501}{2721600} \bigg\} \notag \\
			&\hspace{2.1cm} +\omega_6 \bigg\{-\frac{\rho_{C^3}}{1200}+\rho_{\text{RC}^2}^2 \left(-\frac{3 \rho_{C^3}}{16}-\frac{5 \rho_{C S^2}}{24}+\frac{\rho_{\text{SC}^2}}{4}+\frac{55}{72}\right)+\rho_{C S^2}\left(\frac{\rho_{C^3}}{400}+\left(\frac{569}{43200}-\frac{\rho_{C^3}}{320}\right) \rho_{\text{SC}^2}+\frac{47 \rho_{\text{SC}^2}^2}{17280}-\frac{31}{3600}\right) \notag \\
			&\hspace{3.25cm} +\rho_{\text{RC}^2} \left(\frac{\rho_{C^3}}{40}+\rho_{C S^2} \left(-\frac{3 \rho_{C^3}}{80}+\frac{11 \rho_{\text{SC}^2}}{720}+\frac{47}{360}\right)+\left(\frac{149}{2160}-\frac{\rho_{C^3}}{32}\right) \rho_{\text{SC}^2}-\frac{1}{24} \rho_{C S^2}^2-\frac{\rho_{S^3}}{6}+\frac{\rho_{\text{SC}^2}^2}{24}-\frac{73}{540}\right) \notag \\
			&\hspace{3.25cm} +\left(-\frac{\rho_{C^3}}{768}-\frac{3509}{1036800}\right) \rho_{\text{SC}^2}^2+\left(\frac{\rho_{C^3}}{480}-\frac{2423}{129600}\right) \rho_{\text{SC}^2}+\rho_{\text{RS}^2} \left(\frac{149 \rho_{C S^2}}{600}-\frac{5 \rho_{\text{RC}^2}}{12}-\frac{91 \rho_{S^3}}{900}+\frac{5 \rho_{\text{SC}^2}}{72}+\frac{8}{45}\right) \notag \\
			&\hspace{3.25cm} +\rho_{C S^2}^2 \left(-\frac{3 \rho_{C^3}}{1600}-\frac{7 \rho_{\text{SC}^2}}{7200}+\frac{277}{6000}\right)-\frac{1}{480} \rho_{C S^2}^3+\rho_{S^3} \left(-\frac{187 \rho_{C S^2}}{6000}-\frac{11 \rho_{\text{SC}^2}}{576}+\frac{91}{3600}\right)+\frac{7 \rho_{\text{RS}^2}^2}{90}-\frac{581 \rho_{S^3}^2}{36000}+\frac{\rho_{\text{SC}^2}^3}{576}-\frac{77}{64800} \bigg\} \notag \\
			&\hspace{2.1cm}+\rho_{\text{RS}^2}\left(\frac{\rho_{C S^2}}{120}-\frac{29 \rho_{S^3}}{900}\right)+\frac{1}{300} \rho_{S^3} \rho_{C S^2}-\frac{7 \rho_{\text{RS}^2}^2}{360}-\frac{11 \rho_{S^3}^2}{1125} \Bigg] \, .
		\end{align}
	\end{minipage}
\end{sideways}

\clearpage

\section{B - Table of real fixed points}

In \autoref{tab:FPs} we list all real non-trivial fixed points of cubic gravity in $d=6$ that we found numerically. The fixed points are sorted by the value of $\omega_6$. All values are given with six significant digits, but have been computed with \precision{} digits.

\setlength\LTleft{0pt}
\setlength\LTright{0pt}
\begin{longtable}{@{\extracolsep{\fill}}c|c|c|c|c|c|c|c|c|c@{}}
\caption{List of numerically found real non-trivial fixed points.}\label{tab:FPs}\\ \hline\hline
	   & $\omega_6$ & $\rho_{R^3}$ & $\rho_{RS^2}$ & $\rho_{RC^2}$ & $\rho_{S^3}$ & $\rho_{S^2C}$ & $\rho_{SC^2}$ & $\rho_{C^3}$ & $\rho_{\mathfrak E}$ \\ \hline
	 1 & -193013 & 8.94205$\times 10^6$ & -10.1479 & 3.31281 & -11.0481 & 9.02130 & 2.82862 & -1.19448 & -0.0317721 \\
 2 & -72956.8 & 1.17814$\times 10^6$ & -8.01198 & 2.76911 & -3.32563 & 3.61423 & 3.00481 & -1.07866 & -1.62278 \\
 3 & -6122.19 & -673110 & 3.28258 & -812.330 & -8.20645 & -16.4129 & 31.4925 & -38.2845 & -42.1377 \\
 4 & -3221.30 & -184936 & 1.41157 & -424.788 & -5.33496 & -12.4834 & 21.7531 & -27.0625 & -28.3821 \\
 5 & -2325.80 & -10734.5 & -0.00888476 & -0.0277600 & -4.85041 & 4.66495 & -1.56277 & -0.536414 & 7.34502 \\
 6 & -1200.88 & -162969 & -208.990 & -100.613 & -28.5755 & 4.24945 & 43.6330 & -2.61642 & -62.5065 \\
 7 & -1149.83 & -41595.3 & -222.038 & -122.501 & -31.5582 & 4.19803 & 48.8781 & -2.88143 & -70.5601 \\
 8 & -416.970 & -1401.13 & 18.3844 & -6.48248 & 0.00432010 & -0.261973 & -2.16776 & 1.15683 & 2.39866 \\
 9 & -206.224 & -309.475 & 2.81437 & -1.00590 & -2.24293 & 4.59648 & -4.01964 & -0.491618 & 14.4151 \\
 10 & -188.185 & 261.770 & 2.10233 & -0.711157 & 0.858213 & 1.55323 & 1.02693 & -1.01046 & -3.68782 \\
 11 & -162.864 & -263.782 & 9.19220 & -3.15843 & 0.454393 & 1.42740 & 1.01217 & -1.02747 & 0.644315 \\
 12 & -129.549 & -150.160 & 3.19569 & -1.08909 & 0.813304 & 1.54015 & 1.02584 & -1.01231 & -4.33402 \\
 13 & -112.245 & 416.283 & 28.1215 & -10.5374 & -9.06381 & 4.16724 & 10.4823 & -0.917638 & -10.9390 \\
 14 & -102.409 & -1047.87 & 1.13069 & -10.5645 & -2.60685 & -4.32641 & 9.15766 & -2.69272 & -9.10647 \\
 15 & -95.5302 & 967.972 & 1.15889 & -9.15319 & -2.64416 & -4.45667 & 9.29178 & -2.71932 & -9.94238 \\
 16 & -89.4048 & -80.2302 & -0.253016 & 0.0519267 & -4.81413 & 4.67464 & -1.59817 & -0.539386 & 9.24663 \\
 17 & -85.4731 & 88.0230 & -2.45823 & 0.733060 & -1.71845 & 1.13305 & -0.103701 & -1.04904 & -0.875521 \\
 18 & -82.4166 & -0.218247 & 7.93292 & -2.63067 & -3.17409 & 2.55572 & 2.09762 & -1.04426 & -1.97855 \\
 19 & -80.1527 & 179.391 & 6.18248 & -2.73764 & 1.28566 & 4.35689 & -2.80067 & -0.913076 & 0.198036 \\
 20 & -72.8583 & -90.1926 & 1.10875 & -9.62374 & -2.57669 & -4.22341 & 9.06086 & -2.67552 & -7.67105 \\ \hline
 21 & -68.0026 & -0.332260 & -3.66960 & 1.32861 & -3.28178 & 3.46358 & 2.90327 & -1.07564 & -1.65030 \\
 22 & -60.3627 & -0.100893 & 2.56100 & -0.878393 & 0.783664 & 1.53137 & 1.02191 & -1.01345 & -5.02816 \\
 23 & -53.5525 & 495.460 & -29.5656 & -1.19991 & -11.4422 & 9.26020 & 4.05030 & -1.59712 & -2.54435 \\
 24 & -49.4972 & -1.72110 & 14.4937 & -4.06110 & -4.99413 & 0.623510 & 1.13018 & -1.05829 & -1.48279 \\
 25 & -45.9715 & -195.448 & -0.751719 & -3.81317 & 1.87930 & 3.75860 & -8.85053 & 0.0786952 & 11.6529 \\
 26 & -43.0229 & 178.190 & -0.745473 & -3.12867 & 1.86368 & 3.72737 & -8.78807 & 0.0767323 & 11.5696 \\
 27 & -40.5017 & 281.624 & -21.8741 & -0.925853 & -10.5252 & 7.02740 & 4.89508 & -1.56048 & -2.33566 \\
 28 & -34.2486 & -17.1311 & -0.757258 & -3.61898 & 1.89315 & 3.78629 & -8.90592 & 0.0803631 & 11.7268 \\
 29 & -33.0127 & 22.9983 & -0.295365 & 0.0394296 & -4.74397 & 4.68913 & -1.66597 & -0.543992 & 14.8776 \\
 30 & -32.3232 & -94.8710 & 0.519077 & -2.83434 & -0.232547 & 2.34666 & -3.14428 & -1.52562 & -0.771763 \\
 31 & -31.4163 & 50.0113 & 2.95968 & -1.76140 & -1.41049 & -2.62639 & 4.30133 & 0.222028 & -5.26132 \\
 32 & -30.8219 & 39.3612 & 2.01707 & -1.15947 & 1.42726 & 2.85037 & -7.99747 & 0.584413 & 10.4817 \\
 33 & -26.2399 & -149.151 & 77.8461 & -11.1237 & -7.58295 & 14.8838 & 2.03246 & -0.794779 & -0.798830 \\
 34 & -25.0299 & 23.5236 & 2.77627 & -1.20456 & -2.70708 & -0.738722 & 1.36700 & -1.01717 & -0.838650 \\
 35 & -23.9253 & 45.4471 & 0.618389 & -1.55418 & -1.54597 & -3.09195 & 4.85056 & 0.151875 & -6.61520 \\
 36 & -23.5454 & 30.1350 & 0.482560 & -0.849622 & -0.517688 & -1.02796 & 0.732051 & -1.02677 & -1.11711 \\
 37 & -23.1758 & -1.44813 & 9.21115 & -2.40011 & 1.06928 & -1.89411 & -0.869628 & 3.17072 & -2.21832 \\
 38 & -22.9028 & -41.4341 & 0.620638 & -1.76424 & -1.55160 & -3.10319 & 4.87305 & 0.154351 & -6.64519 \\
 39 & -22.3597 & -6.03914 & 0.621377 & -2.47490 & -0.291910 & 2.27697 & -2.98642 & -1.49849 & -1.06620 \\
 40 & -22.2095 & 34.3116 & 1.90013 & -1.51229 & -0.575901 & 1.93826 & -2.37098 & -1.30112 & -0.728126 \\ \hline
 41 & -21.4963 & -0.314935 & 3.35740 & -1.35526 & -1.04249 & 1.39013 & -1.18427 & -1.15004 & -1.15439 \\
 42 & -18.2294 & -19.8964 & 0.206320 & -0.845435 & -0.515801 & -1.03160 & 0.729869 & -1.03064 & -1.12095 \\
 43 & -17.1276 & -3.08172 & 0.622750 & -1.70395 & -1.55688 & -3.11375 & 4.89417 & 0.156640 & -6.67335 \\
 44 & -15.5428 & -21.7184 & -0.814359 & -0.998000 & -2.36213 & 4.68116 & -3.92656 & -0.532382 & 14.6281 \\
 45 & -15.4839 & -21.3104 & -1.61881 & -0.795576 & 1.76811 & 2.52138 & -8.19208 & 0.788432 & 10.2842 \\
 46 & -15.2871 & 21.4340 & -1.47648 & -0.552400 & 1.73987 & 2.47577 & -8.12437 & 0.815555 & 10.0392 \\
 47 & -15.2060 & -0.191563 & 2.28358 & -0.920899 & -2.60395 & 0.640146 & 1.21746 & -1.02220 & -0.598787 \\
 48 & -14.7329 & 9.74698 & 0.750012 & -0.422632 & -3.35272 & 4.69269 & -3.05081 & -0.499529 & 111.407 \\
 49 & -13.4227 & -0.0160042 & 0.303589 & -0.133482 & -4.17269 & 4.72537 & -2.23420 & -0.536249 & -43.8811 \\
 50 & -13.3666 & -1.03720 & 0.206664 & -0.829382 & -0.516661 & -1.03332 & 0.733309 & -1.03597 & -1.12553 \\
 51 & -12.4490 & -2.34612 & -1.52858 & -0.835619 & 1.79320 & 2.55560 & -8.24337 & 0.768767 & 10.4444 \\
 52 & -12.1300 & -2.03116 & -0.865830 & -0.987965 & -2.41324 & 4.70536 & -3.89195 & -0.541796 & 14.9859 \\
 53 & -11.9955 & -3.11655 & 0.406902 & -0.0559834 & -3.39293 & 3.71062 & 3.02779 & -1.07603 & -1.61241 \\
 54 & -11.9881 & -2.90575 & -4.82875 & 1.62653 & -2.64447 & -1.95653 & 0.716892 & -1.04835 & -5.72683 \\
 55 & -11.4220 & -14.5813 & -6.79329 & -0.119414 & -4.73401 & 4.67583 & -1.92677 & -0.596763 & 24.7833 \\
 56 & -10.4292 & -0.248916 & 1.23256 & -0.691822 & -2.60746 & -0.220529 & 1.30025 & -1.01637 & -0.900132 \\
 57 & -10.1250 & -7.44186 & -6.53023 & -0.111184 & -4.78450 & 4.66511 & -1.87636 & -0.603512 & 17.6887 \\
 58 & -9.77565 & -3.82006 & 0.370390 & -0.0789453 & -10.1363 & 5.96116 & 4.76013 & -1.36909 & -1.96852 \\
 59 & -9.39512 & -0.452532 & 3.90631 & -1.07915 & -4.57259 & 2.71979 & 2.05186 & -1.04551 & -1.57270 \\
 60 & -9.27613 & 4.93566 & 0.179254 & -0.0885014 & -10.1700 & 5.98751 & 4.76625 & -1.37311 & -1.97851 \\ \hline
 61 & -7.48093 & -1.05417 & 9.32845 & -2.82204 & -7.48373 & 5.61771 & 8.13921 & -0.541592 & -0.250805 \\
 62 & -7.20449 & 0.000516166 & 0.284482 & -0.0809885 & -10.0713 & 6.03000 & 4.80741 & -1.37372 & -1.98425 \\
 63 & -7.17059 & -0.438107 & 6.99150 & -2.14209 & -0.183918 & -1.00366 & -2.69281 & -1.46260 & 5.36430 \\
 64 & -7.08654 & 3.88298 & -1.61110 & 0.157932 & -0.785555 & -0.636028 & 0.889065 & -1.00958 & -1.10996 \\
 65 & -6.41347 & 6.25612 & -2.36464 & -0.0968100 & -3.29729 & 4.03480 & 3.33363 & -1.10774 & -1.65666 \\
 66 & -6.24530 & 0.994190 & 2.85318 & -0.858861 & -3.48802 & 0.392699 & 1.16366 & -1.03355 & -1.74873 \\
 67 & -5.35331 & -3.22419 & -2.03284 & -0.120195 & -3.32893 & 4.10070 & 3.37610 & -1.11081 & -1.65829 \\
 68 & -4.63080 & -1.54154 & -1.63353 & 0.0722158 & -0.940365 & -0.399189 & 1.00778 & -1.00699 & -1.12075 \\
 69 & -4.02828 & 1.60298 & -0.416467 & 0.0255380 & -3.45933 & 3.97381 & 3.21981 & -1.08241 & -1.61405 \\
 70 & -3.69645 & 0.0171730 & -1.32088 & 0.305672 & -2.87506 & -0.855325 & 2.17886 & -1.00934 & -1.50883 \\
 71 & -3.64367 & -0.373261 & -1.48679 & 0.0914004 & -0.976633 & -0.345409 & 1.05454 & -1.00613 & -1.12600 \\
 72 & -2.82511 & 0.779750 & -1.63451 & 0.393930 & -3.03169 & -1.71904 & 1.95476 & -1.01418 & -1.26720 \\
 73 & -2.08533 & -0.714779 & 3.68093 & -0.405823 & 1.30448 & -2.10681 & -1.46276 & 3.65330 & -1.85198 \\
 74 & -2.06158 & 0.102649 & -2.15405 & 0.715568 & -2.65392 & -2.40790 & 1.02117 & -1.03913 & -14.0818 \\
 75 & -1.81913 & 0.416168 & -1.57171 & 0.419193 & -2.99951 & -2.20925 & 1.73763 & -1.01897 & -0.804551 \\
 76 & -1.18790 & 0.0412493 & -0.752024 & 0.448916 & 0.741591 & 1.36490 & 1.52690 & -1.04164 & -1.72712 \\
 77 & -1.17935 & 0.0522706 & -1.16565 & 0.521435 & -2.81313 & 2.47513 & 1.98115 & -1.06912 & -2.02384 \\
 78 & -0.972711 & 0.586163 & -2.81734 & 0.275335 & -1.40806 & 3.66123 & 2.62110 & -1.02103 & -2.08539 \\
 79 & -0.728348 & -0.125189 & 0.680386 & 0.115959 & 0.799252 & -1.34212 & 1.59126 & 2.61236 & -0.983215 \\
 80 & -0.716720 & 0.257403 & -3.04491 & 0.835001 & -9.19129 & 4.92986 & 7.42345 & -1.13114 & 4.08435 \\ \hline
 81 & -0.695395 & 0.151261 & -1.37837 & 0.424520 & 0.773378 & 1.99736 & -7.87633 & 1.47237 & -42.1239 \\
 82 & -0.681932 & 0.316942 & -2.87683 & 0.623922 & -8.69917 & 5.46013 & 6.03783 & -1.16896 & 6.05741 \\
 83 & -0.648639 & -0.0805044 & 1.93219 & -0.532314 & 0.413517 & -1.73371 & -2.76579 & 2.28237 & -141.665 \\
 84 & -0.619229 & 0.108926 & -2.08437 & 0.932730 & -2.08943 & 5.20986 & -0.923735 & -0.964488 & 4.75688 \\
 85 & -0.593895 & 0.0197674 & 0.492845 & 0.0491982 & 0.733752 & -1.28735 & 1.54483 & 2.44562 & -1.11247 \\
 86 & -0.509967 & -0.0997698 & 1.37582 & -0.0387213 & -0.640629 & 0.149461 & -0.883793 & -1.11426 & -1.13952 \\
 87 & -0.505503 & -0.0406394 & 0.869039 & -0.0949793 & -0.300559 & 0.741685 & -2.88225 & -1.38801 & -3.88827 \\
 88 & -0.490365 & 0.112826 & -2.22182 & 0.984140 & -1.53457 & 5.57245 & -3.25422 & -0.873620 & 3.66621 \\
 89 & -0.371317 & 0.0470235 & -0.832515 & 0.359425 & -2.51637 & -3.19770 & 1.81211 & -1.04214 & -3.09689 \\
 90 & -0.330702 & -0.0882367 & 2.46775 & -0.768834 & -0.515681 & -5.56856 & 0.482529 & 2.01916 & -71.0488 \\
 91 & -0.323304 & 0.0764663 & -0.511576 & -0.0166054 & -1.42703 & -5.30259 & 4.34914 & 1.02336 & -39.4221 \\
 92 & -0.314644 & -0.00217310 & 0.0391822 & 0.164918 & 1.08615 & 1.78328 & -7.70656 & 1.41541 & 6.35535 \\
 93 & -0.274590 & 0.0257394 & -0.672535 & 0.489740 & -0.0345106 & 0.806863 & 2.31828 & -1.23682 & -2.47615 \\
 94 & -0.257927 & 0.00143843 & -0.00697311 & 0.169315 & 1.02208 & 1.96134 & -7.71736 & 1.27596 & 7.50032 \\
 95 & -0.236522 & 0.0613982 & -1.34032 & 0.777832 & -0.478544 & 1.79392 & 0.840797 & -1.57928 & -0.371445 \\
 96 & -0.211865 & 0.0434316 & -0.776996 & 0.511971 & -1.30192 & 0.640322 & -0.314680 & -1.22540 & -0.847509 \\
 97 & -0.209843 & 0.0576439 & -1.00333 & 0.357363 & 0.839443 & 2.04457 & -7.86977 & 1.52019 & -3.69047 \\
 98 & -0.173923 & 0.0394119 & -0.472857 & 0.285691 & 0.847255 & 1.26594 & 2.67835 & -1.10703 & -1.74378 \\
 99 & -0.157758 & -0.00429563 & -0.0710334 & 0.340195 & 1.05005 & 2.14748 & -7.82572 & 1.08575 & 9.02482 \\
 100 & -0.141831 & 0.0192909 & -0.187453 & -0.0332843 & -1.51921 & -2.16794 & 4.19547 & 0.223039 & -5.44845 \\ \hline
 101 & -0.124047 & 0.0294021 & -0.572720 & 0.388045 & 0.0908647 & 0.266987 & -3.69782 & -1.44453 & -1.53137 \\
 102 & -0.103925 & -0.0122182 & 0.341257 & 0.0389777 & 0.0509618 & -0.604132 & -4.39779 & 1.86022 & 2.13733 \\
 103 & -0.100664 & 0.311608 & -4.84151 & 0.756341 & 8.88184 & 15.2357 & 18.3958 & 4.19020 & -2.03563 \\
 104 & -0.100631 & -0.0427469 & 0.936626 & -0.0345390 & -1.65878 & -2.62319 & 4.52285 & 0.197587 & -6.45338 \\
 105 & -0.0840771 & 0.00724033 & 0.0245818 & -0.120579 & -1.70315 & -5.43340 & 4.53073 & 1.78671 & 6.64720 \\
 106 & -0.0725477 & 0.0195149 & -0.365837 & 0.305184 & 3.96263 & 2.70722 & -14.5106 & 18.7392 & 5.76224 \\
 107 & -0.0687170 & -0.0261214 & 0.572202 & 0.0689466 & -0.749177 & -2.51739 & -0.418827 & -1.18210 & -1.13478 \\
 108 & -0.0501305 & 0.0103088 & -0.197028 & 0.256559 & -0.404122 & -2.20795 & 1.17132 & -1.06389 & -2.74637 \\
 109 & -0.0453856 & -0.160873 & 3.67297 & -1.08726 & -6.48337 & -13.7959 & -1.65558 & 2.75367 & -48.7969 \\
 110 & -0.0317710 & -0.0247962 & 0.581354 & -0.0174946 & -0.367987 & 1.35531 & -3.40530 & 1.41585 & 0.204922 \\
 111 & -0.0272310 & 0.0175498 & -0.373894 & 0.294002 & -0.502894 & -3.88050 & 7.05671 & -4.00018 & 6.15400 \\
 112 & -0.0229536 & -0.0264755 & 0.603179 & 0.140168 & -0.407670 & 1.78076 & -4.90721 & 0.208775 & 0.398989 \\
 113 & -0.0216299 & 0.0225841 & -0.492225 & 0.340760 & 1.33824 & 2.89449 & -8.17891 & 0.502960 & 9.00616 \\
 114 & -0.0157008 & -0.00898911 & 0.205913 & 0.131254 & -0.514782 & -1.02956 & 0.725795 & -1.02431 & -1.11551 \\
 115 & -0.0142219 & -0.00900428 & 0.206216 & 0.0836971 & -0.515540 & -1.03108 & 0.728827 & -1.02902 & -1.11956 \\
 116 & -0.00914804 & -0.00891165 & 0.205888 & 0.0892834 & -0.514511 & -1.02136 & 0.744930 & -1.02596 & -1.11924 \\
 117 & -0.00908644 & -0.0275106 & 0.620058 & -0.0622902 & -1.55015 & -3.10029 & 4.86725 & 0.153717 & -6.63745 \\
 118 & -0.00473799 & -0.0270085 & 0.610686 & -0.0497359 & -1.52669 & -3.05069 & 4.80635 & 0.157951 & -6.49928 \\
 119 & 0.00147288 & 0.00359835 & -0.0847341 & 0.0437821 & 0.223407 & 0.0731520 & -3.06774 & 3.09775 & 2.60356 \\
 120 & 0.00182054 & -0.0374995 & 0.838985 & -0.828975 & -2.12418 & -1.78396 & 12.4361 & 5.87467 & 4.59149 \\ \hline
 121 & 0.00269343 & 0.000331051 & -0.00915760 & 0.224550 & 0.0113301 & 0.0573416 & -2.22605 & 1.48912 & 4.63468 \\
 122 & 0.00307913 & 0.000654459 & -0.0171933 & 0.210229 & 0.0429833 & 0.0859666 & -1.50527 & 0.864610 & 1.85924 \\
 123 & 0.00350302 & -0.0152574 & 0.337614 & -0.161288 & -0.867804 & -0.569524 & 4.83329 & -1.49732 & -4.10997 \\
 124 & 0.00356441 & -0.00113585 & 0.0238920 & 0.224235 & -0.0800039 & -0.108166 & -2.39117 & 1.89275 & 10.1454 \\
 125 & 0.00405023 & 0.000638760 & -0.0170742 & 0.151818 & 0.0426855 & 0.0853710 & -1.50408 & 0.863810 & 1.85765 \\
 126 & 0.00426426 & 0.00116816 & -0.0299028 & 0.209743 & 0.0762615 & 0.217279 & -0.837270 & 0.881027 & 1.75996 \\
 127 & 0.00905142 & 0.0149529 & -0.348847 & 0.104422 & 0.802641 & 0.434371 & -0.0585337 & -1.04924 & -2.29766 \\
 128 & 0.00955430 & 0.000296672 & -0.0188883 & 0.184053 & 0.0472207 & 0.0944413 & -1.52222 & 0.876068 & 1.88183 \\
 129 & 0.0102921 & 0.0162813 & -0.361557 & 0.233919 & 0.902585 & 1.48563 & -1.86005 & -1.01658 & 10.1081 \\
 130 & 0.0107172 & 0.000829027 & -0.0172238 & 0.134098 & 0.0430595 & 0.0861190 & -1.50557 & 0.864815 & 1.85964 \\
 131 & 0.0116327 & 0.000548079 & -0.00484886 & 0.254180 & -0.00566850 & 0.147646 & -2.62505 & 1.30177 & 2.80924 \\
 132 & 0.0117886 & -0.0103209 & 0.222949 & 0.0662111 & -0.585733 & -0.842839 & 2.19095 & 0.759195 & 0.135413 \\
 133 & 0.0185250 & 0.00118406 & -0.0175741 & 0.241656 & 0.0439353 & 0.0878706 & -1.50907 & 0.867171 & 1.86431 \\
 134 & 0.0287373 & -0.000219979 & -0.0194885 & 0.167050 & 0.230839 & -0.273915 & -3.96116 & -1.22974 & 1.49080 \\
 135 & 0.0295578 & 0.0213783 & -0.516490 & 0.0253237 & 0.968642 & -0.953440 & 1.78549 & -0.788310 & -1.11923 \\
 136 & 0.0359359 & 0.0185430 & -0.452002 & 0.325432 & 1.33659 & 2.25747 & -7.46509 & 0.961781 & 7.82408 \\
 137 & 0.0360090 & 0.0125689 & -0.310752 & 0.237617 & 0.955577 & 0.960375 & -0.000336730 & -0.912874 & -5.02676 \\
 138 & 0.0370186 & -0.00582439 & 0.151453 & 0.0303542 & -0.553363 & -0.0256126 & 2.17016 & 0.668741 & 1.47422 \\
 139 & 0.0396187 & -0.0208091 & 0.403273 & 0.0606177 & -1.10014 & -1.55464 & 3.39625 & -1.03143 & -2.85370 \\
 140 & 0.0423319 & -0.0599335 & 1.11549 & 0.0540231 & -2.19878 & -9.22932 & 3.09292 & -3.09101 & -2.95542 \\ \hline
 141 & 0.0532643 & -0.0199841 & 0.328080 & -0.117425 & -0.225916 & -2.91905 & 1.44799 & 0.0677003 & -1.17768 \\
 142 & 0.0571739 & 0.0655554 & -1.50210 & 0.790086 & 2.78133 & 3.72553 & -4.99749 & 2.63405 & 2.73118 \\
 143 & 0.0591459 & 0.0265530 & -0.634945 & 0.0472867 & 1.08921 & -1.24227 & 1.54146 & -0.835957 & -1.13182 \\
 144 & 0.0628796 & 0.00139577 & -0.116492 & 0.208557 & -0.179293 & 0.937003 & 0.725169 & 0.916251 & 3.77396 \\
 145 & 0.0795574 & 0.00382107 & -0.179336 & 0.232856 & 1.33082 & -1.56441 & -9.68091 & 4.36308 & 2.75431 \\
 146 & 0.0809466 & -0.0131903 & 0.159010 & 0.117225 & -0.753314 & -0.975701 & 0.563769 & -1.01804 & -1.04430 \\
 147 & 0.0841316 & 0.0291532 & -0.594762 & 0.209896 & 1.84842 & 0.924189 & -0.330571 & -1.02071 & 10.4144 \\
 148 & 0.0863364 & 0.0209562 & -0.303603 & 0.204342 & 0.552013 & 3.40229 & -1.80317 & -1.70320 & -1.39160 \\
 149 & 0.0881278 & 0.0364155 & -0.778153 & 0.0982690 & 4.02542 & -2.01866 & -16.2988 & 7.88248 & -6.63204 \\
 150 & 0.0983809 & -0.0312208 & 0.492245 & 0.0840312 & -1.61469 & -1.65000 & 3.89269 & 0.0702731 & -3.38130 \\
 151 & 0.101913 & 0.0231827 & -0.804312 & 0.436381 & 2.70597 & 5.87085 & -10.7763 & -0.669247 & 12.1208 \\
 152 & 0.116382 & 0.0237679 & -0.753076 & 0.431230 & 1.23943 & 3.46883 & -7.69810 & 0.189140 & 10.6934 \\
 153 & 0.132758 & 0.0462356 & -1.11830 & 0.531885 & 3.45541 & 10.0038 & -14.4707 & -1.25285 & 14.0141 \\
 154 & 0.134829 & 0.0774796 & -1.29197 & 0.481605 & 6.12933 & 2.41228 & -3.49638 & -2.63111 & -3.53154 \\
 155 & 0.139073 & -0.0709523 & 1.48556 & -0.312451 & -3.78732 & -6.25720 & 13.6615 & -4.29577 & -22.2073 \\
 156 & 0.149552 & 0.0432322 & -1.18027 & 0.566501 & 0.366107 & 5.80562 & -0.383495 & 1.77820 & 17.6571 \\
 157 & 0.154017 & -0.0340534 & 0.329209 & 0.0125896 & -0.719573 & 0.587645 & -0.461562 & 0.134692 & 0.398810 \\
 158 & 0.154434 & 0.0259174 & -0.889070 & 0.657215 & 6.53968 & -12.2651 & -22.4897 & 10.4014 & 10.7727 \\
 159 & 0.179765 & 0.0216261 & -0.313682 & -0.0381656 & 3.81620 & -4.18254 & -10.0785 & -0.235988 & 2.90270 \\
 160 & 0.183297 & 0.0743242 & -2.07384 & 0.635676 & 1.74031 & 3.45834 & -3.60932 & 1.33748 & 0.563105 \\ \hline
 161 & 0.202534 & 0.111804 & -2.77624 & 1.11229 & 2.54781 & 11.6577 & -4.10998 & 2.23549 & 24.2366 \\
 162 & 0.211094 & 0.0267341 & -0.967725 & 0.487986 & -1.43430 & 6.34990 & 4.03099 & 1.61410 & 10.5780 \\
 163 & 0.229166 & 0.0179121 & -0.732758 & 0.446758 & 1.60020 & 3.33097 & -2.29840 & 0.590821 & 3.07323 \\
 164 & 0.241452 & -0.111760 & 2.36724 & -0.671664 & -11.1926 & -5.29813 & -4.64492 & 5.64972 & -39.6576 \\
 165 & 0.254157 & 0.0993810 & -3.15346 & 0.834364 & 1.82080 & 3.04841 & -2.88842 & 1.45999 & -1.53109 \\
 166 & 0.256637 & -0.00966096 & 0.143497 & 0.000942472 & -5.40721 & 10.4263 & 12.8988 & 0.224512 & -2.47992 \\
 167 & 0.266084 & -0.0277747 & 0.0934283 & 0.110467 & -2.09029 & -1.59091 & 3.11967 & -1.03192 & -2.79604 \\
 168 & 0.268221 & 0.0330412 & 0.181982 & 0.359652 & -1.34976 & 0.817410 & -0.0550249 & 0.0251512 & 0.243563 \\
 169 & 0.323587 & -0.0105886 & -0.0819538 & 0.0875005 & -1.30774 & 3.89413 & -0.335938 & -0.824944 & -0.689247 \\
 170 & 0.326532 & 0.0403574 & -0.373328 & 0.125613 & -1.00034 & 4.06058 & -1.25986 & -0.738850 & -0.203170 \\
 171 & 0.370720 & 0.0295919 & -0.855070 & 0.277679 & -1.60662 & 1.07084 & 1.25866 & -0.974713 & -1.74830 \\
 172 & 0.425066 & 0.00459618 & -0.835185 & 0.412156 & -1.88631 & 1.25006 & 1.05648 & -0.935125 & -3.86362 \\
 173 & 0.426855 & 0.00981768 & -0.355476 & 0.214372 & 1.66632 & 2.05944 & -7.77900 & 1.08703 & 8.08043 \\
 174 & 0.430720 & -0.00771256 & -1.02856 & 0.524661 & -0.512128 & 4.11494 & -5.44471 & -0.219919 & 8.31975 \\
 175 & 0.446963 & 0.0270889 & -1.50905 & 0.661978 & -1.96537 & 4.81618 & -0.991544 & -0.591085 & 6.35231 \\
 176 & 0.453761 & 0.580768 & 3.16259 & -1.25585 & -117.488 & 16.7079 & 89.0631 & -3.28545 & -36.7192 \\
 177 & 0.498709 & -0.0364383 & -0.630162 & 0.405864 & 1.53342 & 3.32096 & -2.06943 & 0.382635 & 1.75704 \\
 178 & 0.573045 & 0.0458461 & -0.369939 & 0.205405 & 1.67395 & 2.08267 & -7.80785 & 1.07268 & 8.21865 \\
 179 & 0.593470 & 0.127470 & -0.112888 & 0.481223 & -2.48080 & 1.43056 & -1.74523 & 0.259151 & 0.667738 \\
 180 & 0.614194 & -0.185084 & -1.07203 & 0.570411 & 25.5890 & -8.47996 & -19.3966 & 2.19048 & 7.85974 \\ \hline
 181 & 0.802966 & -0.00485694 & -0.0986808 & 0.176899 & 1.28854 & 3.79436 & -2.11413 & -1.41296 & -1.08140 \\
 182 & 0.812245 & 0.124455 & -0.190323 & 0.198509 & 1.25443 & 3.80434 & -2.16185 & -1.39273 & -0.991965 \\
 183 & 0.818700 & 0.223043 & -0.604839 & 0.319010 & 2.34296 & 8.25923 & -6.41299 & -1.43020 & 2.81380 \\
 184 & 0.841049 & 0.0149544 & -0.503858 & 0.301928 & 3.12996 & 11.6083 & -8.74287 & -1.69410 & 4.63302 \\
 185 & 0.867570 & 0.00555251 & 0.533771 & -0.209644 & 7.43766 & -9.51036 & -1.50380 & 0.443362 & 0.229148 \\
 186 & 0.907986 & -0.00617261 & -0.0771975 & 0.169139 & 0.963283 & 2.26073 & -0.293865 & -1.05238 & -0.991178 \\
 187 & 0.911855 & -0.362664 & -1.01777 & 0.565387 & 20.9070 & -7.13971 & -21.3301 & 2.43800 & 10.7569 \\
 188 & 0.914635 & -0.133284 & -0.203145 & 0.235867 & 0.862247 & 1.81082 & 0.418402 & -1.01553 & -0.612780 \\
 189 & 0.947661 & -0.00375397 & -0.151473 & 0.201786 & 0.866469 & 1.70234 & 0.649671 & -1.01294 & -0.280996 \\
 190 & 0.982249 & 0.110952 & -0.0952661 & 0.167786 & 0.960262 & 2.28981 & -0.340908 & -1.05475 & -0.998510 \\
 191 & 1.04317 & 0.0504782 & -1.57834 & 0.677018 & -2.47431 & 4.38238 & -3.66539 & -0.321573 & 10.3506 \\
 192 & 1.09709 & 0.0378136 & -1.32937 & 0.489223 & 18.5730 & -5.78930 & -21.3592 & 2.27163 & 11.8689 \\
 193 & 1.16488 & 0.100241 & -0.140915 & 0.192205 & 0.868542 & 1.65428 & 0.763086 & -1.01172 & 0.216769 \\
 194 & 1.18951 & -0.230664 & -0.0471879 & 0.171308 & 0.968556 & 2.30601 & -0.363168 & -1.05552 & -1.01131 \\
 195 & 1.19642 & -0.306520 & 0.0875763 & 0.129585 & 1.31544 & 3.81180 & -2.14600 & -1.43533 & -1.14453 \\
 196 & 1.49840 & 0.0859141 & -1.78786 & 0.516331 & -1.84046 & -0.521218 & -0.352053 & -1.03290 & -0.525569 \\
 197 & 1.65257 & 0.885713 & -2.37662 & 0.672286 & 20.4801 & -6.55826 & -23.2362 & 2.58461 & 13.4087 \\
 198 & 1.66218 & 0.225203 & -3.52940 & 1.26079 & -1.99938 & 1.87209 & -1.87524 & -0.942037 & 272.547 \\
 199 & 1.92456 & 0.805863 & -0.332359 & 0.00232766 & -14.9507 & 12.4762 & 3.24217 & -0.981710 & -1.25068 \\
 200 & 2.17756 & -0.735791 & -0.817595 & 0.754690 & 1.76614 & 1.77184 & -7.64343 & 1.31087 & 6.48020 \\ \hline
 201 & 2.21245 & 0.918634 & 0.899404 & -0.396561 & 9.72236 & -11.6709 & -3.41916 & 0.539710 & 0.773181 \\
 202 & 3.00147 & -0.711504 & 1.08171 & -0.408598 & -3.80565 & 2.48258 & -1.50388 & -0.0555702 & 0.226396 \\
 203 & 3.22074 & -0.00951398 & -0.0730736 & 0.0288803 & -12.4357 & 10.1041 & 2.66971 & -1.04092 & -1.21594 \\
 204 & 4.56201 & -0.391077 & -1.23741 & 1.23614 & 1.87458 & 1.43971 & -7.26435 & 1.54155 & 3.57824 \\
 205 & 5.96612 & 0.172966 & -5.51555 & 1.95917 & -3.48624 & 4.15974 & -2.21318 & -0.612223 & 13.8939 \\
 206 & 6.36453 & -4.75737 & -2.24198 & 1.57454 & 1.60536 & 2.96398 & -2.07424 & 0.520689 & 2.03915 \\
 207 & 7.31216 & -0.286254 & -2.06859 & 1.38783 & 0.455586 & -1.35497 & 1.74612 & 1.73097 & -1.46501 \\
 208 & 8.78744 & -1.30235 & -1.47934 & 1.80743 & 1.82614 & 1.08311 & -6.61450 & 1.78644 & -0.978971 \\
 209 & 8.86649 & -4.12920 & -4.31560 & 1.69472 & -3.68064 & 4.46388 & -2.44057 & -0.534464 & 15.3832 \\
 210 & 9.38400 & -0.663706 & -6.91909 & 2.35107 & -4.24486 & 4.12109 & -1.72135 & -0.535470 & 4.56527 \\
 211 & 9.79294 & -2.63289 & -0.0168526 & 1.82697 & 0.0421314 & 0.0842628 & -1.50186 & 0.862324 & 1.85469 \\
 212 & 10.0134 & 14.1174 & -0.0175450 & 1.54688 & 0.0438624 & 0.0877249 & -1.50878 & 0.866975 & 1.86392 \\
 213 & 10.0774 & -0.0267913 & 0.0652418 & -0.0148874 & -11.2600 & 9.14102 & 2.63503 & -1.13476 & -1.03095 \\
 214 & 14.2023 & -24.9158 & -0.0171590 & 2.01558 & 0.0428976 & 0.0857952 & -1.50492 & 0.864380 & 1.85878 \\
 215 & 16.0438 & -12.7517 & -5.76144 & 2.48240 & -3.48007 & 5.11920 & 3.73157 & -1.10955 & -1.59927 \\
 216 & 16.6718 & -29.4930 & -1.43808 & 2.30791 & 1.69780 & 0.822095 & -6.03143 & 1.92081 & -3.52683 \\
 217 & 17.1202 & -19.7928 & -2.36232 & 1.77682 & 0.590627 & -1.28306 & 1.62703 & 2.02712 & -1.29335 \\
 218 & 41.5075 & -19.5171 & 6.14033 & -1.84834 & -3.38484 & 4.15509 & 3.60201 & -1.11678 & -1.55936 \\
 219 & 219.366 & -387.731 & 3.94230 & -0.836875 & 0.672070 & -1.21253 & 1.53419 & 2.23735 & -1.28609 \\
 220 & 440.239 & -75.5208 & -15.7661 & 5.44960 & 0.390974 & -1.18162 & 1.67642 & 1.62957 & -2.11610 \\ \hline\hline
\end{longtable}

\end{document}